\documentclass[final,12pt,authoryear]{elsarticle}
\usepackage[margin=2.5cm]{geometry}% by courtesy of Mico

\usepackage{graphicx}
\usepackage{textcomp}

\usepackage{amsfonts,amssymb,amsmath,amsthm,autobreak,bm,cancel,dsfont,mathtools,mathbbol,mathrsfs,siunitx,upgreek}

\usepackage{tabularx} % Add this to the preamble
\usepackage{caption} % Add this package
\usepackage{booktabs} % For better looking tables
\usepackage{float}
\usepackage[hidelinks,colorlinks=true,urlcolor=blue,linkcolor=black,citecolor=blue]{hyperref}
\usepackage[mathlines]{lineno}

\usepackage{cite}  % Optional: to manage multiple citations better
\usepackage{xcolor}
\usepackage{makecell}
\usepackage{appendix}
\usepackage{parskip}                    % default spacing around environments

\usepackage{multirow}  % For multirow cells
\usepackage{adjustbox} % To adjust table size
\begin{document}

\begin{frontmatter}



% \title{Harnessing the Stretch-Bend Synergy of Tension-Induced Soft Stress in Liquid Crystal Elastomers and Viscoelastic Bending for Enhanced Energy Dissipation} %% Article title

\title{Tension-Induced Soft Stress and Viscoelastic Bending in Liquid Crystal Elastomers for Enhanced Energy Dissipation} %% Article title

% OR
%\title{Enhancing Energy Dissipation in Architected Structures by Leveraging the Tension-Induced Soft Stress Response of Liquid Crystal Elastomers} %% Article title


% \author[inst1]{Beijun Shen}

\author[inst1]{Beijun Shen\textsuperscript{\dag}\corref{cor1}}

% \author[inst1]{Beijun Shen\textsuperscript{\dag,*}}

\fntext[dag]{Beijun Shen has moved to Columbia University}
\ead{bs3660@columbia.edu}

\author[inst1]{Yuefeng Jiang}
\author[inst2]{Christopher M. Yakacki}
\author[inst1,inst3]{Sung Hoon Kang\corref{cor1}}
\ead{shkang2024@kaist.ac.kr}

\cortext[cor1]{Corresponding authors}

\author[inst1]{Thao D. Nguyen}
% \ead{vicky.nguyen@jhu.edu}

\address[inst1]{Department of Mechanical Engineering, Johns Hopkins University, Baltimore, Maryland 21218, United States of America}
\address[inst2]{Department of Mechanical Engineering, University of Colorado, Denver, CO 80204, USA}
\address[inst3]{Department of Materials Science and Engineering, Korea Advanced Institute of Science and Technology, Daejeon 34141, Republic of Korea}

%% Abstract
\begin{abstract}
Architected materials that exploit buckling instabilities to reversibly trap energy have been shown to be effective for impact protection. The energy-absorbing capabilities of these architected materials can be enhanced further by incorporating viscoelastic material behavior into the buckling elements using liquid crystal elastomers (LCE). In addition to conventional viscoelastic behavior, LCEs also exhibit a highly dissipative rate-dependent soft stress response from mesogen rotation under a mechanical load. However, the buckling elements cannot take advantage of this dissipation mechanism because buckling occurs at strains below the threshold for mesogen rotation.  In this study, we investigate tension-induced soft stress behavior as an additional dissipation mechanism in horizontal members of lattice structures composed of tilted LCE beams under compression. Viscoelastic properties of LCEs with two crosslinking densities were characterized experimentally, and a nonlinear viscoelastic model was implemented in Abaqus/Standard as a user-defined element to simulate finite-strain behavior of monodomain LCEs, including soft stress response. Simulations and experiments revealed a non-monotonic dependence of energy dissipation on the thickness ratio between horizontal and tilted LCE members. Optimized structures with stretchable horizontal bars dissipated 2–3 times more energy than rigid-bar counterparts by balancing tension-driven soft stress with viscoelastic beam bending. These findings demonstrate a new design strategy for LCE-based architected materials to enhance energy dissipation.

\end{abstract}

%% Keywords
\begin{keyword}
Liquid crystal elastomers \sep viscoelasticity \sep soft elasticity \sep buckling \sep energy dissipation


% MSC codes here, in the form: \MSC code \sep code
% or \MSC[2008] code \sep code (2000 is the default)

\end{keyword}

\end{frontmatter}



% \linenumbers
\section{Introduction} 

Architected materials have gained increasing attention over the past few decades for applications in impact mitigation and vibration isolation. Their energy absorption performance is governed by the interplay between structural topology and the intrinsic properties of the constituent materials. While early studies primarily focused on harnessing geometric instabilities such as elastic buckling to absorb energy through large recoverable deformations \citep{Shan2015MultistableEnergy, Haghpanah2016MultistableMaterials, Restrepo2015PhaseMaterials, Bertoldi2017HarnessingMaterials, Frenzel2016TailoredAbsorbers, Qiu2004AMechanism, Chen2021ReusableColumns, Liu2019ArchitectedDissipation}, more recent efforts have demonstrated that energy absorption can be dramatically enhanced by integrating these instabilities with material dissipation mechanisms, such as plasticity \citep{Yao2022TailoringDissipations, Liu2024HarnessingAbsorption} and viscoelasticity \citep{Jeon2022SynergisticElastomersb, Yao2024TunableRates}. Among these mechanisms, viscoelasticity is especially attractive due to its rate-dependent behavior, which provides enhanced protection at higher strain rates, and its fully recoverable nature, which enables repeated use without permanent damage.

In previous work, we developed a pseudo-bistable lattice structure composed of repeating unit cells with tilted beams fabricated from liquid crystal elastomers (LCEs) sandwiched between rigid horizontal supports \citep{Jeon2022SynergisticElastomersb}. LCEs are crosslinked polymer networks containing stiff, rod-like molecular motifs called mesogens, which develop orientational order in response to mechanical loading. This coupling between mesogen rotation and mechanical deformation gives rise to a rate-dependent semi-soft stress response, where the material undergoes large deformations with minimal increase in stress \citep{Azoug2016ViscoelasticityElastomers, MartinLinares2020TheElastomer}, and enhanced dissipation relative to conventional elastomers \citep{Clarke2001SoftElastomers, Merkel2019MechanicalLoading}. The viscoelastic behavior of the LCE beams caused the energy absorption of each unit cell in the lattice to increase according to a power-law dependence on strain rate, resulting in a tenfold greater energy absorption density compared to an otherwise identical unit cell constructed from PDMS under intermediate impact strain rates.  Furthermore, the progressive buckling of unit cells within the lattice structure induced repeated load-unload cycles within adjacent unit cells that amplified the viscoelastic dissipation beyond the expected additive contributions of the individual components.

Various groups have also exploited the superior viscoelastic dissipation to enhance energy absorption in LCE architected materials. Traugutt et al. \citep{Traugutt2020Liquid-Crystal-Elastomer-BasedPrinting} reported that polydomain LCE structures printed by digital light processing had 12 times greater rate-dependence and up to 27 times greater energy dissipation compared to those printed from a commercially available photocurable elastomer resin. Telles et al. \citep{Telles2025ArchitectedAbsorption} showed that well-aligned LCE lattices printed by direct ink writing exhibited 18 times higher energy absorption than their silicon counterparts under high impact strain rates (up to \(10^3/\text{s}\)). Other relevant studies on LCE lattices are found in Bischoff et al. \citep{Bischoff2024MonodomainDamping} and Song et al. \citep{Song2024OnLattices}. 
The prevailing designs of architected materials absorb energy through viscoelastic bending of beam elements, which cannot take advantage of the highly dissipative soft stress response unique to LCEs.  The maximum strains induced by beam bending are typically limited to 30–40\%, which is lower than the typical threshold for mesogen rotation. Because mesogen rotation is largely irreversible, the soft-stress mechanisms present significant potential for further enhancing energy dissipation in LCE-based architected systems.

Mistry et al.~\citep{Mistry2021SoftElastomers} reported plateau-like soft stress enabling impact absorption at nearly constant stress, and Shaha et al.~\citep{Shaha2020BiocompatibleDisc} observed stronger soft stress under transverse tension than longitudinal compression, likely due to increased mesogen reorientation. Luo et al.~\citep{Luo20213DInsultb} investigated the soft stress mechanism in 3D-printed LCE architected materials and found that increased lattice connectivity in polydomain LCE foams enhanced energy dissipation, which they attributed to greater axial strain and possible mesogen rotation, though this was not directly measured. Previous studies on bi-material lattices incorporated rigid inclined struts connected to highly extensible transverse elements via pin joints~\citep{Ye2023MultimaterialMetamaterials, Ruschel2020ALattices}. Energy dissipation in these systems arose solely from the axial stretching of the transverse elements under vertical compression, while the inclined beams rotated freely and did not contribute to dissipation. Viscoelastic bending or buckling of the inclined beams was not utilized to enhance energy absorption. Furthermore, the transverse elements were made of conventional elastomers, such as thermoplastic polyurethane (TPU), rather than LCEs. Consequently, these designs did not leverage the unique soft stress response of LCEs, which enables energy dissipation through tension-induced mesogen rotation.

In this study, we investigate incorporating the rate-dependent soft-stress deformation mechanism into an LCE architected material through a tension element.
Building on our previous design \citep{Jeon2022SynergisticElastomersb}, we replace the rigid horizontal member with a stretchable LCE bar. Under compression, the tilted beams undergo bending while the horizontal bar stretches, thereby activating the soft-stress response through mesogen rotation. However, the lateral expansion of the unit cell from the stretched bar reduces the tilt angle of the beams, diminishing bending and its associated dissipation. This tradeoff creates a nonlinear interaction between viscoelastic bending and soft-stress deformation mechanisms. We systematically examine the competition between viscoelastic bending and soft-stress tensile dissipation mechanisms using finite element simulations and experiments, varying both the LCE material properties and the thickness ratio between the horizontal and tilted members. A nonlinear viscoelastic model was implemented in Abaqus/Standard as a user-defined element (UEL) to capture the large deformation behavior of monodomain LCEs and to quantify the respective contributions of the different deformation mechanisms to the unit cell's overall energy dissipation. Our results reveal a non-monotonic dependence of energy dissipation on the thickness ratio, with maximum absorption achieved at an intermediate value. Structures with the optimal thickness ratio between stretching and bending elements dissipated two to three times more energy than those with rigid horizontal bars, thereby demonstrating the synergistic benefit of combining viscoelastic bending and tension-induced soft-stress responses.

\section{Methods}
\label{Sec:Section_2}

\subsection{Experimental Methods}

\subsubsection{LCE material preparation}
Impressio (Denver, CO, USA) synthesized and supplied two sets of LCEs with different crosslinking densities, formulated to replicate the chemical, mechanical, and morphological behavior reported in prior studies \citep{Yakacki2015TailorableReaction, Saed2017Thiol-acrylateStrain}. The fabrication process involved a two-stage thiol-acrylate Michael addition, mechanical alignment, and subsequent photopolymerization to produce main-chain monodomain LCEs. Polydomain LCEs were fabricated using the same procedure but without mechanical alignment. The precursor mixture consisted of 2,2'-(Ethylenedioxy)diethanethiol (EDDET), Pentaerythritol tetrakis(3-mercaptopropionate) (PETMP), and 1,4-Bis-[4-(3-acryloyloxypropoxy)benzoyloxy]-2-methylbenzene (RM257). Two batches of polydomain LCEs with varying crosslinking densities were prepared. To obtain monodomain LCEs, approximately 100\% strain was applied to half of the polydomain samples from each batch, aligning the mesogens through a nematic phase transition via director rotation. The resulting polydomain sheets had an average thickness of approximately 0.93 mm, and the monodomain sheets measured approximately 0.64 mm. The specific catalysts, initiators, and processing conditions remain proprietary to Impressio.

\subsubsection{Unit cell structure preparation}
\label{sec:StructurePreparation}

\begin{figure}[H]
   \centering     
   \includegraphics[width=0.95\textwidth]{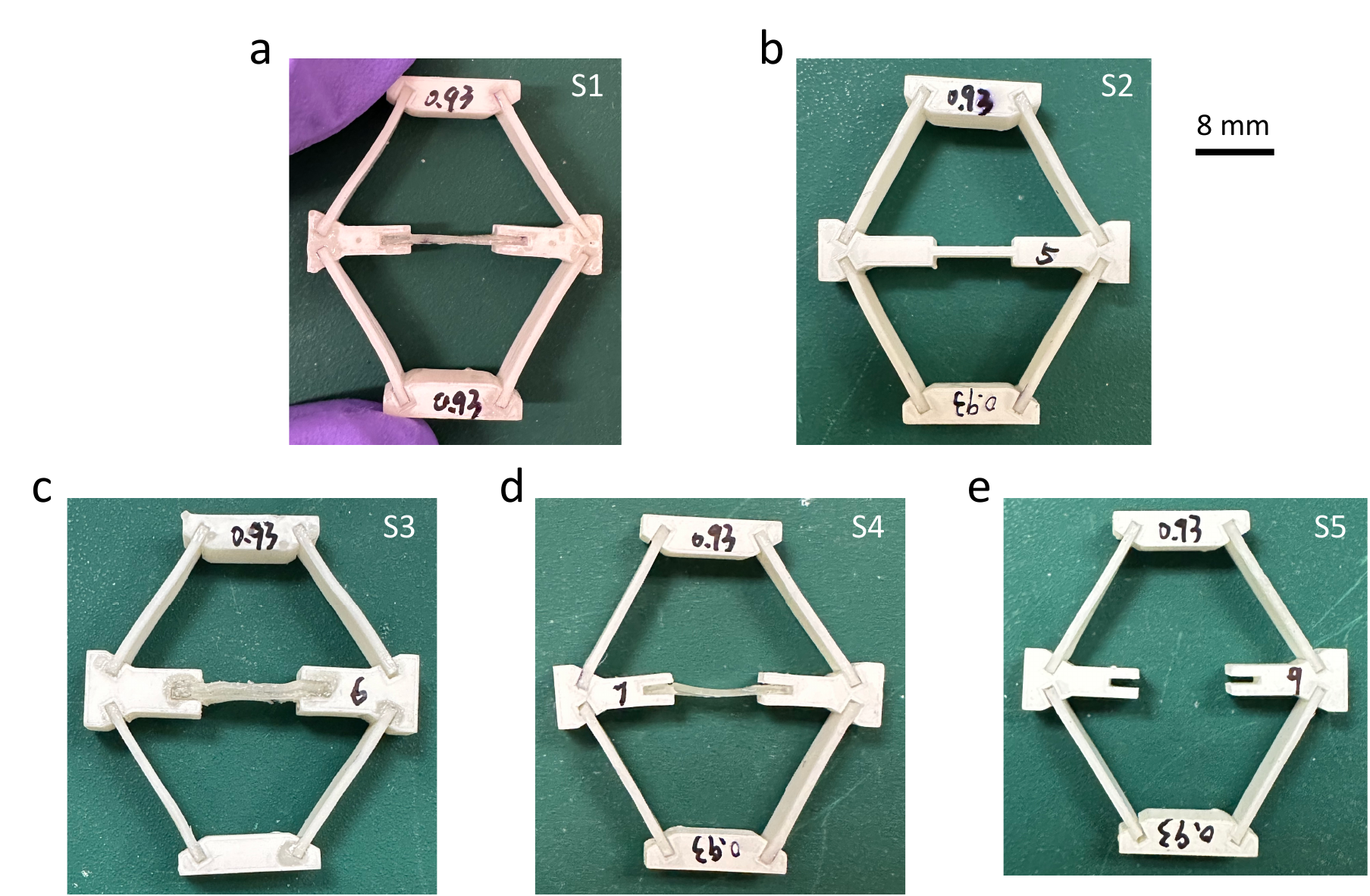}
   \caption{Experimental unit cell structures. All tilted beams are made of polydomain LCE with higher crosslinking density and are fixed at a 60$^\circ$ angle relative to the horizontal member. The horizontal member varies across designs: (a) Unit cell 1 — perpendicular LCE with lower crosslinking density, 0.64~mm thickness; (b) Unit cell 2 — rigid horizontal bar, 0.64~mm thickness; (c) Unit cell 3 — perpendicular LCE with lower crosslinking density, 1.28~mm thickness; (d) Unit cell 4 — perpendicular LCE with higher crosslinking density, 0.64~mm thickness; (e) Unit cell 5 — no horizontal bar.}
   \label{fig:Fig_1}
\end{figure}

Unit cell structures were fabricated using LCEs with different crosslinking densities (Figure~\ref{fig:Fig_1}). LCE strips were cut from prepared sheets to specified dimensions. Two types of strips were used: polydomain LCEs and monodomain LCEs, the latter featuring an initial director perpendicular to the longitudinal axis (hereafter referred to as perpendicular LCEs). Each unit cell consisted of four identical tilted LCE beams arranged in a hexagonal configuration and a horizontal transverse member, with variations in material and thickness across designs. The in-plane thickness of the horizontal member is defined as \( t_h \), and the in-plane thickness of the tilted beams is defined as \( t_t \); these are key parameters in the subsequent parametric study. This layout enabled a systematic investigation of how material and geometric variations influence mechanical response and energy dissipation. All components were assembled using 3D-printed rigid joints made from Rigid 10K Resin (FormLab 2B, Formlabs Inc., Somerville, MA, USA) and bonded with Gorilla Super Glue (Cincinnati, OH, USA). The adhesive was cured overnight before testing.

Each unit cell had an in-plane width of approximately 30~mm. The total height was approximately 35~mm for the structures shown in Figure~\ref{fig:Fig_1}a, b, d, and e, and about 37~mm for the one in Figure~\ref{fig:Fig_1}c. The out-of-plane thickness was 5~mm for all designs. Each tilted LCE beam had an effective length of 13.4~mm, a thickness of 0.93~mm, and was oriented at a 60$^\circ$ angle relative to the horizontal member. All tilted beams were composed of polydomain LCE with higher crosslinking density. The 60$^\circ$ inclination was chosen to balance competing effects. Larger angles reduce the horizontal force component available to stretch the horizontal member, whereas smaller angles limit vertical displacement under compression, thereby diminishing both beam bending and horizontal stretching.

The horizontal member, when present, had an effective length of 8~mm and an out-of-plane thickness of 5~mm. The thickness and material of the horizontal bar were varied across five unit cell configurations. Unit cell 1 included a perpendicular LCE bar with lower crosslinking density and a thickness of 0.64 mm. Unit cell 2 used a rigid bar of the same thickness made from Rigid 10K Resin to constrain deformation. Unit cell 3 maintained the same LCE formulation as unit cell 1 but doubled the thickness to 1.28~mm by bonding two identical strips using 3M\texttrademark{} VHB\texttrademark{} Tape F9460PC. Unit cell 4 employed a perpendicular LCE bar with higher crosslinking density and a thickness of 0.64~mm, enabling direct comparison with unit cell 1 to isolate the effect of crosslinking density. Unit cell 5 omitted the horizontal member entirely, serving as a control to assess its contribution to mechanical performance. In all cases where LCE horizontal bars were used, the initial director was perpendicular to the longitudinal direction and oriented into the page in Figure~\ref{fig:Fig_1}. Table~\ref{tab:SampleKey} summarizes the configurations.

To eliminate residual strain from fabrication and handling, all unit cells underwent thermal resetting prior to mechanical testing. The structures were heated on a hot plate at 125$^\circ$C -- above the LCE nematic-to-isotropic transition temperature for 5 minutes and then cooled to room temperature for at least 20 minutes. This treatment ensured a uniform initial state prior to mechanical testing. Following thermal resetting, the unit cells were subjected to vertical compression to evaluate their mechanical response and energy dissipation.

\begin{table}[ht]
    \centering
    \caption{Summary of unit cell configurations. Tilted beams are composed of polydomain LCEs with higher crosslinking density (CLD) and a fixed in-plane thickness \( t_t = 0.93 \,\text{mm} \). The horizontal member varies in material and thickness \( t_h \) as listed.}
    \small
    \begin{tabular}{@{}ccc@{}}
        \toprule \toprule
        \textbf{Unit cell}    & \textbf{Horizontal member} & \textbf{Thickness $\bm{t_h}$ (mm)}  \\
        \toprule \toprule
        Unit cell 1    & Lower CLD, $\perp$    & 0.64  \\ 
        Unit cell 2    & Rigid                 & 0.64  \\ 
        Unit cell 3    & Lower CLD, $\perp$    & 1.28  \\ 
        Unit cell 4    & Higher CLD, $\perp$   & 0.64  \\ 
        Unit cell 5    & None                  & –  \\ 
        \bottomrule
    \end{tabular}    
    \label{tab:SampleKey}
\end{table}

\subsubsection{Dynamic mechanical analysis (DMA) tests}
\label{sec:DMA}
Dynamic mechanical analysis was conducted using a DMA 850 (TA Instruments, New Castle, DE, USA) equipped with a liquid nitrogen cooling system to characterize the viscoelastic properties of both monodomain and polydomain LCEs. LCE sheets were cut into strips measuring 20~mm in length and approximately 9~mm in width. Each specimen was mounted in tension clamps with an active length of 5–10~mm and subjected to sinusoidal dynamic strain with an amplitude of 0.1\%. Oscillatory temperature and frequency sweeps were performed to evaluate the viscoelastic response. The temperature was increased from $-40~^\circ$C to 150~$^\circ$C in 5~$^\circ$C increments. At each temperature, frequency sweeps ranging from 0.02~Hz to 20~Hz were conducted to measure the storage modulus ($E'$) and loss modulus ($E''$). Samples were held isothermally for 5 minutes before data acquisition at each temperature to ensure thermal equilibrium. Before testing, all specimens were thermally reset by heating at 125~$^\circ$C for 5 minutes, then cooling to room temperature for at least 20 minutes to eliminate residual strains and establish consistent initial conditions. Master curves for storage modulus and tan$\delta$ were constructed by horizontally shifting the frequency sweep data using the time–temperature superposition (TTS) principle, referenced to 25~$^\circ$C \citep{Ferry1980ViscoelasticPolymers}.

\subsubsection{Uniaxial tension load-unload tests}
\label{sec:uniaxial}
The large-strain viscoelastic behavior of LCEs was characterized through uniaxial tension load–unload tests. Monodomain LCE strips were cut to 50~mm in length, 10~mm in width, and 0.64~mm in thickness, with the long axis either parallel or perpendicular to the pre-stretch direction. Approximately 10~mm at each end was reserved for gripping, leaving a 10~mm gauge length at the center, which was marked with a black permanent marker for strain tracking.
Quasi-static tests were performed under displacement control using an MTS Insight 5 universal testing machine equipped with a 500~N load cell (MTS Systems, Eden Prairie, MN, USA). Specimens were mounted between tensile grips, and a preload of 0.06~N was applied to remove slack. This value was determined based on observed load fluctuations at zero force.
For each crosslinking density, three specimens were tested in both parallel and perpendicular orientations to assess variability. Tests were conducted at four strain rates (0.01\%/s, 0.1\%/s, 1\%/s, and 10\%/s) to capture rate-dependent behavior. Prior to testing, all samples were thermally reset by heating at 125~$^\circ$C for 5 minutes, followed by cooling at room temperature for at least 20 minutes to ensure consistent initial conditions.
Target strains varied with crosslinking density and orientation. Perpendicular specimens were stretched to capture both the soft stress plateau associated with mesogen rotation and the strain-stiffening regime due to polymer chain alignment. To avoid fracture, specimens with lower and higher crosslinking densities were stretched to 800\% and 700\% engineering strain, respectively. Parallel LCEs were stretched to 80\% and 70\% strain for the lower and higher crosslinking density, respectively. Engineering stress was calculated as the reaction force divided by the initial cross-sectional area, and engineering strain as the applied displacement divided by the gauge length. Young’s modulus was determined from a linear fit to the initial 5\% strain range. For perpendicular LCEs, peak stress was defined as the maximum stress preceding strain softening. Hysteresis was quantified by numerically integrating the enclosed area within the stress–strain loop using the trapezoidal rule.

\subsubsection{Compression tests of unit cell structures}
\label{sec:CompressionTests}
Quasi-static uniaxial compression tests were conducted to evaluate the mechanical response and energy absorption of the unit cell structures and to validate finite element simulations. Tests were performed under displacement control using an MTS Insight 5 universal testing machine equipped with a 100~N load cell (MTS Systems, Eden Prairie, MN, USA).
Each unit cell was positioned between two rigid cylindrical plugs, each 12~mm in diameter and 15~mm in height, fabricated from Rigid 10K Resin (FormLab 2B, Formlabs Inc., Somerville, MA, USA). The lower plug was affixed to the bottom compression plate using double-sided tape to prevent sliding, while the upper plug was mounted to the upper compression plate to ensure symmetric loading. This plug–unit cell–plug configuration prevented mechanical interference between the unit cell and the platens during compression.

Prior to testing, each unit cell was thermally reset following the procedure in Section~\ref{sec:StructurePreparation} to eliminate residual strains and ensure consistent initial conditions. The upper platen was manually lowered until the upper plug contacted the top surface of the unit cell without inducing deformation. The load cell was then zeroed, and compression was initiated. Tests were performed at three effective strain rates: 1\%/s, 10\%/s, and 50\%/s. Each unit cell was compressed to a total displacement of 20~mm. The corresponding displacement rates were calculated by multiplying the strain rate by the target displacement to maintain consistency across tests. Crosshead displacement was recorded as the applied displacement, and the load cell measured the reaction force. 

Each configuration was tested three times at each strain rate to evaluate repeatability and variability. Energy absorption was computed by numerically integrating the area under the load-displacement curve using the trapezoidal rule. Raw data collected at 50\%/s were smoothed to reduce noise.

\subsection{Modeling Methods}

\subsubsection{Finite deformation quasi-linear viscoelastic model}
\label{sec:QLV}

Compression of the unit cell structures induces bending in the tilted beams, generating tensile strains below 40\%, which are below the threshold for semi-soft stress associated with mesogen rotation in perpendicular-loaded LCEs. In this moderate strain regime, the material exhibits classical viscoelastic behavior. We modeled this response using the finite-deformation viscoelastic formulation available in Abaqus/Standard~\citep{dassaultsystemesSIMULIAUserAssistance2023}.

This viscoelastic framework assumes that the strain and time dependence of the stress response are separable. The total stress is decomposed into a time-independent hyperelastic component and a time-dependent viscoelastic contribution governed by a dimensionless shear relaxation function \( g(t) \), defined as:
\begin{equation}
    g(t) = \frac{\mu(t)}{\mu_0},
\end{equation}
where \( \mu(t) \) is the shear modulus at time \( t \), and \( \mu_0 = \mu(0) \) is the instantaneous shear modulus. By definition, \( g(0) = 1 \) and \( g(\infty) = \mu_\infty / \mu_0 \), where \( \mu_\infty \) denotes the long-term (equilibrium) shear modulus. The relaxation function is approximated using a Prony series:
\begin{equation}
    g(t) = g_\infty + \sum_{i=1}^{N} g_i \, e^{-t/\tau_i},
\end{equation}
where \( g_\infty = \mu_\infty/\mu_0 \), \( \tau_i \) are characteristic relaxation times, and the coefficients satisfy \( \sum_{i=1}^{N} g_i = 1 - g_\infty \). These parameters were calibrated by fitting the DMA-derived master curves of the storage modulus under uniaxial tension (Section~\ref{sec:DMA}). The full calibration procedure and final parameter values are detailed in Appendix~\ref{app:ParameterForTiltedBeams}.

The dimensionless relaxation function \( g(t) \) enters the hereditary integral formulation of the Cauchy stress:
\begin{equation}
\begin{split}
   \boldsymbol{\sigma}(\mathbf{F}, t) &= \boldsymbol{\sigma}^D\left(\overline{\mathbf{F}}, t\right) + \boldsymbol{\sigma}^H(J), \\
   \boldsymbol{\sigma}^D(\mathbf{F}, t) &= \boldsymbol{\sigma}_0^D(\mathbf{F}(t)) + \mathrm{dev} \left[ \int_{0}^{t} \dot{g}(t') \, \overline{\mathbf{F}}_t^{-1}(t - t') \cdot \boldsymbol{\sigma}_0^D(\mathbf{F}(t - t')) \cdot \overline{\mathbf{F}}_t^{-\top}(t - t') \, dt' \right],
\end{split}
\label{eq:sigvisco}
\end{equation}
where \( \boldsymbol{\sigma} \) is the total Cauchy stress, decomposed into deviatoric and volumetric components, \( \boldsymbol{\sigma}^D \) and \( \boldsymbol{\sigma}^H \), respectively. The quantity \( \boldsymbol{\sigma}_0^D \) denotes the instantaneous deviatoric Cauchy stress, \( \mathbf{F} \) is the deformation gradient, \( J = \det(\mathbf{F}) \) is the volume ratio, and \( \overline{\mathbf{F}} = J^{-1/3} \mathbf{F} \) is the distortional (isochoric) part of \( \mathbf{F} \). The tensor \( \overline{\mathbf{F}}_t(t - t') \) denotes the relative distortional deformation gradient from time \( t - t' \) to \( t \). The deviatoric projection is defined as \( \mathrm{dev}(\mathbf{A}) = \mathbf{A} - \frac{1}{3} \, \text{tr}(\mathbf{A}) \, \mathbf{I} \). For simplicity, the volumetric response \( \boldsymbol{\sigma}^H \) is assumed time-independent.

To define the rate-independent elastic response, we employed a hyperelastic model. In Abaqus, this model may be applied to either the instantaneous stress \( \boldsymbol{\sigma}_0^D \) or the equilibrium stress \( \boldsymbol{\sigma}_\infty^D \). We chose the latter, since the long-time, high-temperature equilibrium state is clearly identifiable in the DMA-measured storage modulus. The deviatoric equilibrium Cauchy stress is computed from the hyperelastic strain energy density:
\begin{equation}
    \boldsymbol{\sigma}_\infty^D = \frac{2}{J} \, \mathrm{dev}\left[\left( \frac{\partial W}{\partial\overline{I}_1} + \overline{I}_1 \frac{\partial W}{\partial\overline{I}_2} \right)\overline{\mathbf{B}} -  \frac{\partial W}{\partial\overline{I}_2}\overline{\mathbf{B}}^2 \right],
\end{equation}
where \( \overline{\mathbf{B}} = \overline{\mathbf{F}} \, \overline{\mathbf{F}}{^\top} \) is the deviatoric left Cauchy–Green tensor, and \( \overline{I}_1 \), \( \overline{I}_2 \) are its first and second invariants.

For simplicity, we adopted the built-in compressible neo-Hookean model in Abaqus to define the hyperelastic strain energy density:
\begin{equation}
    W = C_1 (\overline{I}_1 - 3) + \frac{1}{D_1} (J - 1)^2,
\end{equation}
where \( 2C_1 = \mu_\infty \) and \( D_1 = 2/K \), with \( \mu_\infty \) denoting the equilibrium shear modulus and \( K \) the bulk modulus. The equilibrium shear modulus is related to the instantaneous modulus by:
\begin{equation}
    \mu_\infty = \mu_0 \left( 1 - \sum_{i=1}^{N} g_i \right).
\end{equation}

We estimated \( \mu_\infty = 0.666 \, \text{MPa} \) from the low-frequency (rubbery plateau) storage modulus measured via DMA for the polydomain LCE with higher crosslinking density. The bulk modulus was specified as \( K = 8733.6 \, \text{MPa} \), calculated from the high-frequency (glassy) storage modulus \( E^\mathrm{neq} \) using
\begin{equation}
    K = \frac{E^\mathrm{neq}}{3(1 - 2\nu_g)},
\end{equation}
where \( \nu_g = 0.35 \) is the assumed Poisson’s ratio in the glassy regime. This physically motivated choice yields a bulk modulus over five orders of magnitude greater than \(\mu_\infty\), ensuring a nearly incompressible response.

\subsubsection{Finite Deformation Viscoelastic Micropolar Model for Monodomain LCEs}
\label{sec:WangModel}

The finite deformation micropolar viscoelastic theory developed by \citet{Wang2022AElastomers} was used to model the anisotropic, large-deformation behavior of monodomain LCEs.  We briefly summarize the constitutive model below. The theory conceptualizes a monodomain as a continuum with a director field $\bm{d}$  that can rotate independently of the deformation gradient $\mathbf{F}$.  The director describes the mean preferred orientation of the mesogens. The degree of alignment is described by the order parameter, $Q$, which is assumed to be constant and independent of deformation. The network deformation tensor is defined as, $\mathbf{F}_N = \bm l^{-\frac{1}{2}} \mathbf{F} \bm l_0^{\frac{1}{2}}$, to represent the deformation of the polymer network relative to the deformation caused by director rotation, and the current $\bm l$ and reference $\bm l_0$ step length tensors are defined as, 
\begin{align}
    \bm{l} &=(1-Q) \left( \mathbf{I} - \bm{d} \otimes \bm{d} \right) + (1+2Q)\bm{d} \otimes \bm{d}, \\
    \bm{l}_0 &= (1-Q) \left( \mathbf{I} - \bm{d}^0 \otimes \bm{d}^0 \right) + (1+2Q)\bm{d}^0 \otimes \bm{d}^0,
\end{align}
where $\bm{d}^0$ is the reference director orientation.

To describe the viscoelastic deformation of the material, a multiplicative decomposition was assumed for the deformation gradient into elastic and viscous parts, $\mathbf{F} = \mathbf{F}^\mathrm{e} \mathbf{F}^\mathrm{v}$. This allows the elastic network deformation to be defined as, $\mathbf{F}^\mathrm{e}_N = \bm l^{-\frac{1}{2}} \mathbf{F}^\mathrm{e}  \bm l_0^{\frac{1}{2}}$,

The following free energy density was assumed for the viscoelastic stress response of the network:  
\begin{equation}
    \Psi( \mathbf{F}_N,\mathbf{F}^\mathrm{e}_N, J,\bm{d}) = \overline{\Psi}^\mathrm{eq}( \mathbf{F}_N) + \overline{\Psi}^\mathrm{neq}(\mathbf{F}^\mathrm{e}_N) + \frac{\kappa}{2} (J - 1)^2 + \frac{\gamma}{4} (\bm{d} \cdot \bm{d} - 1)^2,
    \label{eq_Psi}
\end{equation}
where $\kappa$ is the bulk modulus of the LCE material, was set to be 1000 times the shear modulus at equilibrium to approximate an incompressible deformation response, and $\gamma$ is a Lagrange multiplier constraining $\bm{d}\cdot\bm{d} =1$. The neo-Gent model was introduced by \citet{Wang2022AElastomers} for the equilibrium and non-equilibrium free energy densities to describe the strain stiffening stress response of the polymer network at large deformation.
The individual components are:
\begin{equation}
\begin{aligned}
    \overline{\Psi}^{\mathrm{eq}} &= -\frac{\mu^{\mathrm{eq}} }{2} I_m \ln\left(1-\frac{I^N - 3}{I_m}\right) - \mu^{\mathrm{eq}} \ln J, \\
    \overline{\Psi}^{\mathrm{neq}} &= -\frac{\mu^{\mathrm{neq}} }{2} I_m \ln\left(1-\frac{I^{N\mathrm{e}} - 3}{I_m}\right) - \mu^{\mathrm{neq}} \ln J^{\mathrm{e}}, 
 \end{aligned}
\label{eq_Psi_specify}
\end{equation}
where $J = \mathrm{det}\left(\mathbf{F}\right)$, $J^{\mathrm{e}} = \mathrm{det}\left(\mathbf{F}^\mathrm{e}\right)$,  $I^N$ and $I^{N\mathrm{e}}$ are the first invariants of $\mathbf{C}_N = \mathbf{F}^{\top}_N \mathbf{F}_N$ and $\mathbf{C}^\mathrm{e}_N = \mathbf{F}^\mathrm{e\top}_N \mathbf{F}^\mathrm{e}_N$, $\mu^\mathrm{eq}$ and  $\mu^\mathrm{neq}$ are the equilibrium and nonequilbrium shear moduli, and $I_m$ is the strain stiffening parameter.   Neglecting the effects of director gradients, the constitutive equations for the stress response, evolution of the viscous deformation gradient, and evolution of the director field are defined from the free energy density as follows:
\begin{align}
    \boldsymbol{\upsigma} &= \frac{1}{J} \text{sym} \left( \frac{\partial \Psi}{\partial \mathbf{F}} \mathbf{F}^{\top} + \frac{\partial \Psi}{\partial \mathbf{F}^\mathrm{e}} \mathbf{F}^{\mathrm{e}{\top}} \right) %= \boldsymbol{\upsigma}_{eq} + \boldsymbol{\upsigma}_{neq}, 
    \label{eq_sigma} \\
    \mathbf{L}^\mathrm{v} &= \frac{1}{\eta_N} \mathbf{F}^{\mathrm{e}{\top}} \frac{\partial \Psi}{\partial \mathbf{F}^\mathrm{e}}, \label{eq_L^v} \\
    \dot{\bm{d}} -\mathbf{W} \bm{d} &= -\frac{1}{J \eta_D} \left[ \frac{\partial \Psi}{\partial \bm{d}} - \left( \frac{\partial \Psi}{\partial \bm{d}} \cdot \bm{d} \right) \bm{d} \right]. \label{eq_d_circle}
\end{align}
Here, $\mathbf{L}^\mathrm{v} = \dot{\mathbf{F}}^\mathrm{v} \mathbf{F}^{\mathrm{v}^{-1}}$ is the viscous rate of deformation tensor, $\mathbf{W}$ is the spin tensor,  $\eta_D$ and $\eta_N$ are the viscosities associated with director rotation and viscous deformation of the polymer network, respectively.

Substituting equations \eqref{eq_Psi} and \eqref{eq_Psi_specify} into \eqref{eq_sigma} gives the Cauchy stress:
\begin{align}
    \boldsymbol{\upsigma} &= \boldsymbol{\upsigma}^\mathrm{eq} + \boldsymbol{\upsigma}^\mathrm{neq},  \\%= \frac{1}{J} \text{sym} \left( \frac{\partial (\overline{\Psi}^{eq} + \Psi^b)}{\partial \mathbf{F}} \mathbf{F}^{\top} \right) + \frac{1}{J} \text{sym} \left( \frac{\partial \overline{\Psi}^{neq}}{\partial \mathbf{F}^e} {\mathbf{F}^e}^{\top} \right), \notag \\
    \text{where} \\
    \boldsymbol{\upsigma}^\mathrm{eq} &= \frac{1}{J} \left( \frac{\mu^\mathrm{eq}}{1 - Q} \frac{I_m}{I_m - (I^N - 3)} \right) \left[ \mathbf{F} \bm{l}_0 \mathbf{F}^{\top} - \frac{3Q}{1 + 2Q} \text{sym} (\bm{d} \otimes \mathbf{F} \bm{l}_0 \mathbf{F}^{\top} \bm{d}) \right] - \left[ \frac{\mu^\mathrm{eq}}{J} - \kappa (J - 1) \right] \mathbf{I}, \notag \\
    \boldsymbol{\upsigma}^\mathrm{neq} &= \frac{1}{J} \left( \frac{\mu^\mathrm{neq}}{1 - Q} \frac{I_m} {I_m - (I^{N\mathrm{e}} - 3)} \right) \left[ \mathbf{F}^\mathrm{e} \bm{l}_0 {\mathbf{F}^\mathrm{e} }^{\top} - \frac{3Q}{1 + 2Q} \text{sym} (\bm{d} \otimes \mathbf{F}^\mathrm{e} \bm{l}_0 {\mathbf{F}^\mathrm{e} }^{\top} \bm{d}) \right] - \frac{\mu^\mathrm{neq} }{J} \mathbf{I}.
\label{eq:cauchy_stress}
\end{align}
Similarly, substituting equations \eqref{eq_Psi} and \eqref{eq_Psi_specify} into \eqref{eq_L^v} gives the viscoelastic flow rule:
\begin{equation}
    \mathbf{L}^\mathrm{v} = \frac{\mu^\mathrm{neq} }{\eta_N} \left( \frac{I_m}{I_m - (I^{N\mathrm{e} } - 3)} \mathbf{F}^{\mathrm{e} {\top}} \bm{l}^{-1} \mathbf{F}^\mathrm{e}  \bm{l}_0 - \mathbf{I} \right).
\end{equation}

For director rotation, substituting equations \eqref{eq_Psi} and \eqref{eq_Psi_specify} into \eqref{eq_d_circle} gives:
\begin{equation}
\begin{aligned}
    \dot{\bm{d}} %&= \mathbf{W}\bm{d} -\frac1{J\eta_D}  \left[ \frac{\partial\Psi}{\partial\bm{d}}-\left(\frac{\partial\Psi}{\partial\bm{d}}\cdot\bm{d}\right)\bm{d} \right] \\ 
    %
%    &= \mathbf{W}\bm{d} -\frac1{J\eta_D} \underbrace{ \left[ \frac{\partial \overline{\Psi}^{eq}}{\partial\bm{d}}-\left(\frac{\partial \overline{\Psi}^{eq}}{\partial\bm{d}}\cdot\bm{d}\right)\bm{d} \right]}_\textbf{$\bm{m}^{eq}$} -\frac1{J\eta_D} \underbrace{\left[\frac{\partial \overline{\Psi}^{neq}}{\partial\bm{d}}-\left(\frac{\partial \overline{\Psi}^{neq}}{\partial\bm{d}}\cdot\bm{d}\right)\bm{d}\right]}_\textbf{$\bm{m}^{neq}$} \\
    %
%    &\quad  - \frac1{J\eta_D} \left[ \frac{\partial \Psi^d} {\partial\bm{d}}-\left(\frac{\partial \Psi^d}{\partial\bm{d}}\cdot\bm{d}\right) \bm{d} \right] \\
    %
    &= \mathbf{W}\bm{d} -\frac1{J\eta_D} (\bm{m}^\mathrm{eq} + \bm{m}^\mathrm{neq})  + \frac{\gamma}{J\eta_D} (\bm{d} \cdot \bm{d} - 1)^2 \bm{d},
\end{aligned}
\label{eq:d_dot_penalty}
\end{equation}
where, 
\begin{equation}
\begin{aligned}
\bm{m}^\mathrm{eq}&=-\mu^\mathrm{eq}\frac{3Q}{(1-Q)(1+2Q)}\frac{I_m}{I_m-(I^N-3)}\left[\mathbf{F}\bm{l}_0\mathbf{F}^{\top}\cdot\bm{d}-(\bm{d}\cdot\mathbf{F}\bm{l}_0\mathbf{F}^{\top}\cdot\bm{d})\bm{d}\right]\\
\bm{m}^\mathrm{neq}&=-\mu^\mathrm{neq}\frac{3Q}{(1-Q)(1+2Q)}\frac{I_m}{I_m-(I^{N\mathrm{e}}-3)}\left[\mathbf{F}^\mathrm{e}\bm{l}_0{\mathbf{F}^\mathrm{e}}^{\top}\cdot\bm{d}-(\bm{d}\cdot\mathbf{F}^\mathrm{e}\bm{l}_0{\mathbf{F}^\mathrm{e}}^{\top}\cdot\bm{d})\bm{d}\right],
\end{aligned}
\end{equation}
%and
%\begin{equation}
%\frac{\partial \Psi^d}{\partial \bm{d}} = \gamma (\bm{d} \cdot \bm{d} - 1) \bm{d}.
%\end{equation}

The evolution of director viscosity is: 
\begin{equation}
\eta_D = \eta_{D_0} \text{exp}(-\frac{\sqrt{\boldsymbol{\upsigma} :\boldsymbol{\upsigma}}}{k_S}).
\end{equation}

The dissipation power density for the model is given by:
\begin{equation}
\mathcal{D}_{\text{int}} = \eta_N \frac{\|\mathbf{L}^\mathrm{v}\|^2}{J} + \eta_D \|\overset{\circ}{\bm{d}}\|^2,
\end{equation}
where the first term represents network dissipation and the second term represents dissipation from director rotation.
The model was implemented as a UEL subroutine \citep{YourGitHubRepo} in Abaqus/Standard for finite element simulations of LCE structures.

The model includes eight parameters: $Q$, $\mu^\mathrm{eq}$, $\mu^\mathrm{neq}$, $\eta_{N}$, $\eta_{D_0}$, $k_S$, $I_m$, and $\theta_0$. These parameters were calibrated using the rate-dependent uniaxial stress response for the two sets of monodomain LCEs with different crosslinking densities, stretched perpendicular to the initial director, as discussed in Section~\ref{sec:uniaxial}. The fitting process considered only the first 400\% engineering strain, covering the semi-soft stress response.  A preliminary study showed that the horizontal LCE bar of the unit cell structures did not stretch beyond 400\% strain. % This calibration established the basis for analyzing more complex structures in Sections~\ref{Sec:Section_3}-\ref{Sec:Section_5}, where the LCE tensile members remained within this stretch limit.
The parameter calibration process using a material point program and the resulting table of parameters are described in Appendix~\ref{app:MaterialPoint}.

\subsubsection{Finite element simulation of uniaxial tension tests}
\label{uniaxialFEA}

The material point computations described in Appendix~\ref{app:MaterialPoint} neglect the influence of boundary-induced inhomogeneities near the grips, which can affect the mechanical response in uniaxial tension~\citep{Chehade2024FiniteElastomers, shen2025dissertation}. We performed full-scale finite element simulations of uniaxial tension tests for LCE strips with different crosslinking densities to account for these effects and validate the nonlinear viscoelastic material model under large deformations. A user-defined element (UEL) implementation of the nonlinear viscoelastic model developed by Wang et al.~\citep{Wang2022AElastomers} was used (see Section~\ref{sec:WangModel}). Model parameters were assigned accordingly: parameters in Table~\ref{tab:Model_SoftPerpLCE} for the less crosslinked network and Table~\ref{tab:Model_StiffPerpLCE} for the more crosslinked network.

The model geometry represented a rectangular strip measuring 30~mm in length ($x$), 6.1~mm in width ($y$), and 0.63~mm in thickness ($z$), discretized using 24,000 eight-node hexahedral elements, each with eight integration points. The mesh resolution was 0.3~mm in $x$, approximately 0.2~mm in $y$, and 0.08~mm in $z$. Boundary conditions were applied by fully fixing the left surface in all directions. The right surface was constrained in the $y$ and $z$ directions and subjected to a prescribed displacement in $x$. A displacement of 120~mm, corresponding to 400\% engineering strain, was applied at an effective strain rate of 1\%/s, followed by unloading at the same rate.

\begin{figure}[H]
    \centering
    \includegraphics[width=0.85\textwidth]{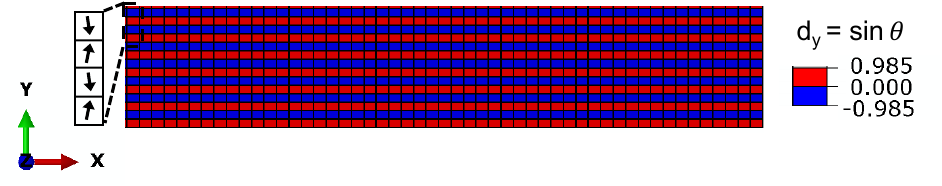}
    \caption{Finite element mesh with an alternating initial director pattern. The color scale represents the $y$-component of the initial director field, $d_y = \sin\theta$, where $\theta$ is the initial director angle relative to the $x$-axis. Alternating strips are assigned $\theta = \pm 80^\circ$, corresponding to $d_y = \pm 0.985$.}
    \label{fig:Fig_2}
\end{figure}

To suppress macroscopic shear deformation from director rotation near the grips and improve numerical stability, the initial director was assigned an alternating pattern along the $y$-direction~\citep{shen2025dissertation, Chehade2024FiniteElastomers}. Figure~\ref{fig:Fig_2} illustrates the initial director field of the more crosslinked network. Adjacent rows of elements were assigned initial director angles of $\pm 80^\circ$ relative to the $x$-axis. This alternating pattern reduced net shear without suppressing local director reorientation and allowed better convergence. The chosen angle matched the experimentally observed peak stress and is discussed further in Appendix~\ref{app:MaterialPoint}.  During simulation, the applied displacement was recorded as the $x$-displacement of the right surface. The total reaction force was computed by summing the $x$-direction reaction forces at all nodes on that surface. Engineering stress was calculated as the total force divided by the undeformed cross-sectional area, and engineering strain was defined as the applied displacement divided by the gauge length.

\subsubsection{Finite element simulation of unit cell structures}
\label{sec:FEA_unitcells}

\begin{figure}[H]
    \centering     
    \includegraphics[width=0.922\textwidth]{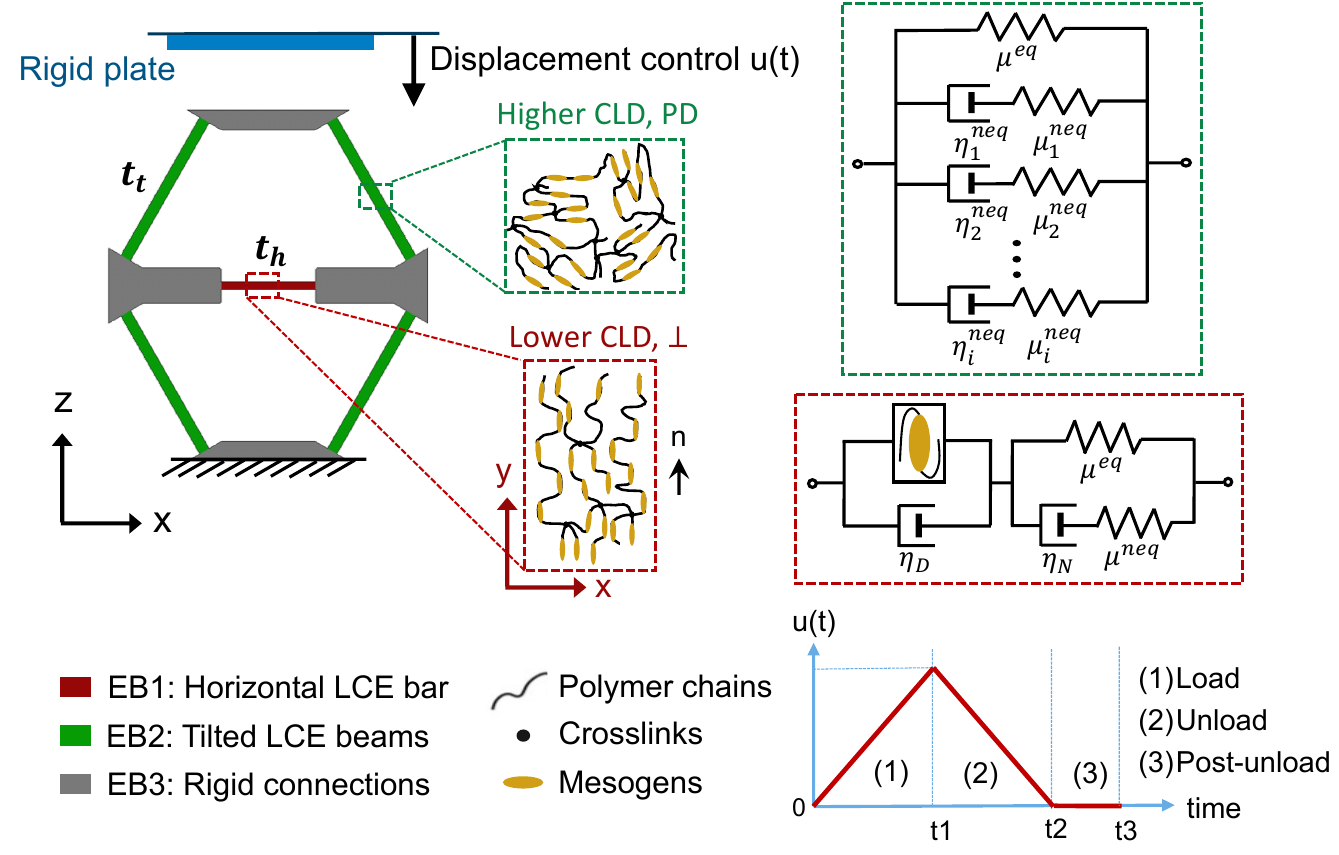}
    \caption{Schematic of the finite element model and boundary conditions. The unit cell comprises a horizontal LCE bar (block 1, red), tilted LCE beams (block 2, green), and rigid joints (block 3, grey). A prescribed displacement \( u(t) \) is applied through a rigid plate to simulate loading, unloading, and post-unload holding. Geometric parameters are labeled: \( t_h \) denotes the in-plane thickness of the horizontal bar, and \( t_t \) denotes the in-plane thickness of the tilted beams. Insets illustrate representative molecular configurations of polydomain (PD) LCEs with higher crosslinking density (CLD) used in the tilted beams, and a perpendicular (\(\perp\)) LCE with lower CLD used in the horizontal bar by default.}
    \label{fig:Fig_3}
\end{figure}

We performed finite element simulations to analyze the mechanical response and energy dissipation of unit cell structures with varying horizontal member thicknesses and material properties. The unit cell geometry matched the experimental configuration (Section~\ref{sec:StructurePreparation}), consisting of four tilted LCE beams inclined at 60°, a horizontal transverse member, and rigid connectors (Figure~\ref{fig:Fig_3}).

The horizontal LCE bar (element block 1, red) was modeled using fully integrated eight-node hexahedral user-defined elements (UELs). The default material parameters corresponded to perpendicular LCEs with lower crosslinking density (Table~\ref{tab:Model_SoftPerpLCE}); additional simulations used higher crosslinking density (Table~\ref{tab:Model_StiffPerpLCE}). The nonlinear viscoelastic constitutive model included dissipation from mesogen rotation and polymer network viscoelasticity. Details of the micropolar viscoelastic model and parameter determination are provided in Section~\ref{sec:WangModel} and Appendix~\ref{app:MaterialPoint}, respectively.

The tilted LCE beams (element block 2, green) were modeled using fully integrated eight-node hybrid elements (C3D8H, Abaqus/Standard), incorporating the built-in finite-deformation viscoelastic model described in Section~\ref{sec:QLV}. Material parameters for polydomain LCEs with higher crosslinking density are listed in Table~\ref{tab:Parameters_StiffPD}. Additional simulations with alternative beam orientations (parallel or perpendicular monodomain LCEs) were also conducted using relaxation spectra shown in Figure~\ref{fig:Fig_S9}b.

Rigid joints (element block 3, grey) were assigned linear elastic properties ($E = 10^4$ MPa, $\nu = 0.4$) consistent with the experimental resin (Rigid 10K, Formlabs Inc., Somerville, MA, USA) and meshed using fully integrated eight-node hexahedral elements (C3D8, Abaqus/Standard).

\begin{figure}[H]
    \centering         
    \includegraphics[width=0.5\textwidth]{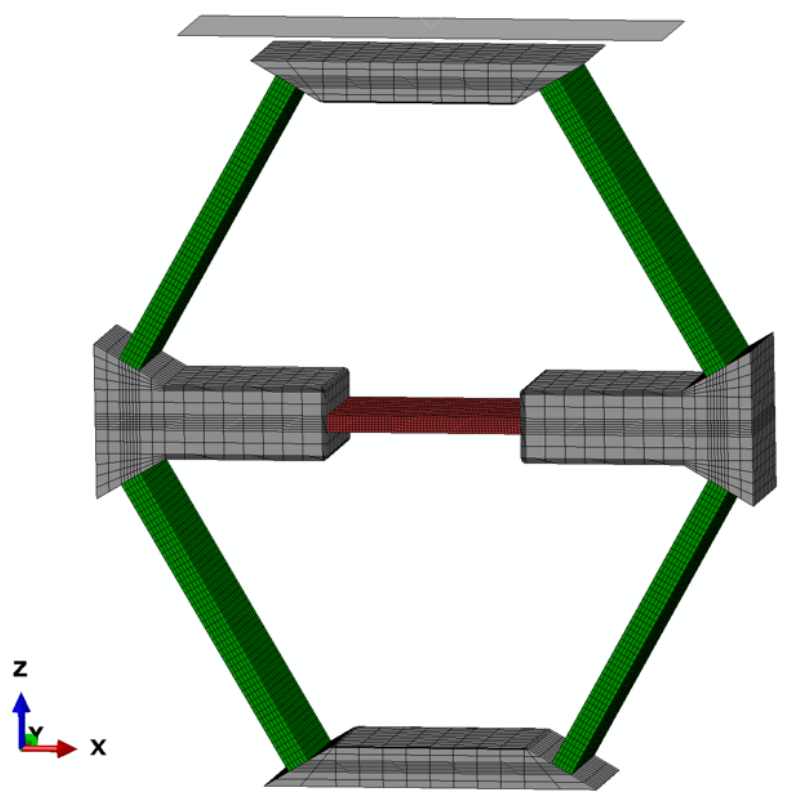}
    \caption{Finite element mesh of the unit cell structure.}
    \label{fig:Fig_4}
\end{figure}

The full finite element mesh contained 31,441 eight-node elements, including 4320 UEL elements in the horizontal bar and 3600 C3D8H elements per tilted beam (Figure~\ref{fig:Fig_4}). For the horizontal bar, element sizes were approximately 0.13~mm ($x$), 0.42~mm ($y$), and 0.11~mm ($z$). For tilted beams, element sizes were approximately 0.268~mm along the beam axis, 0.155~mm transversely, and 0.42~mm through out-of-plane thickness ($y$). Element sizes were selected based on a convergence study to ensure numerical accuracy. Boundary conditions are shown in Figure~\ref{fig:Fig_3}. The bottom surface was fixed, and lateral surfaces were constrained
% to be continued

% to be continued
to allow symmetric $x$-direction displacements. 
Out-of-plane ($y$-direction) motion of the rigid joints was restricted. A prescribed vertical displacement was applied to a rigid plate in three stages: loading, unloading, and post-unloading hold. Loading and unloading rates matched those used in experiments (Section~\ref{sec:CompressionTests}), and the top plate was held in its initial position for 30 seconds following unloading to observe the viscoelastic recovery of the structures. Vertical reaction force and displacement of the top surface were recorded throughout.

For the tilted LCE beams modeled with built-in elements, total elastic strain energy, internal energy, and viscoelastic dissipation were extracted directly from Abaqus output. For the horizontal LCE bar modeled with UEL elements, a custom post-processing algorithm was developed to decompose the total energy into contributions from elastic storage, mesogen rotation, and polymer network dissipation. Total energy absorption was calculated by numerically integrating the load-displacement loop curve using the trapezoidal rule.

\section{Results and Discussions}

\subsection{Finite deformation viscoelastic behavior of LCEs under uniaxial tension}

\begin{figure}[H]
    \centering       
    \includegraphics[width=0.95\textwidth]{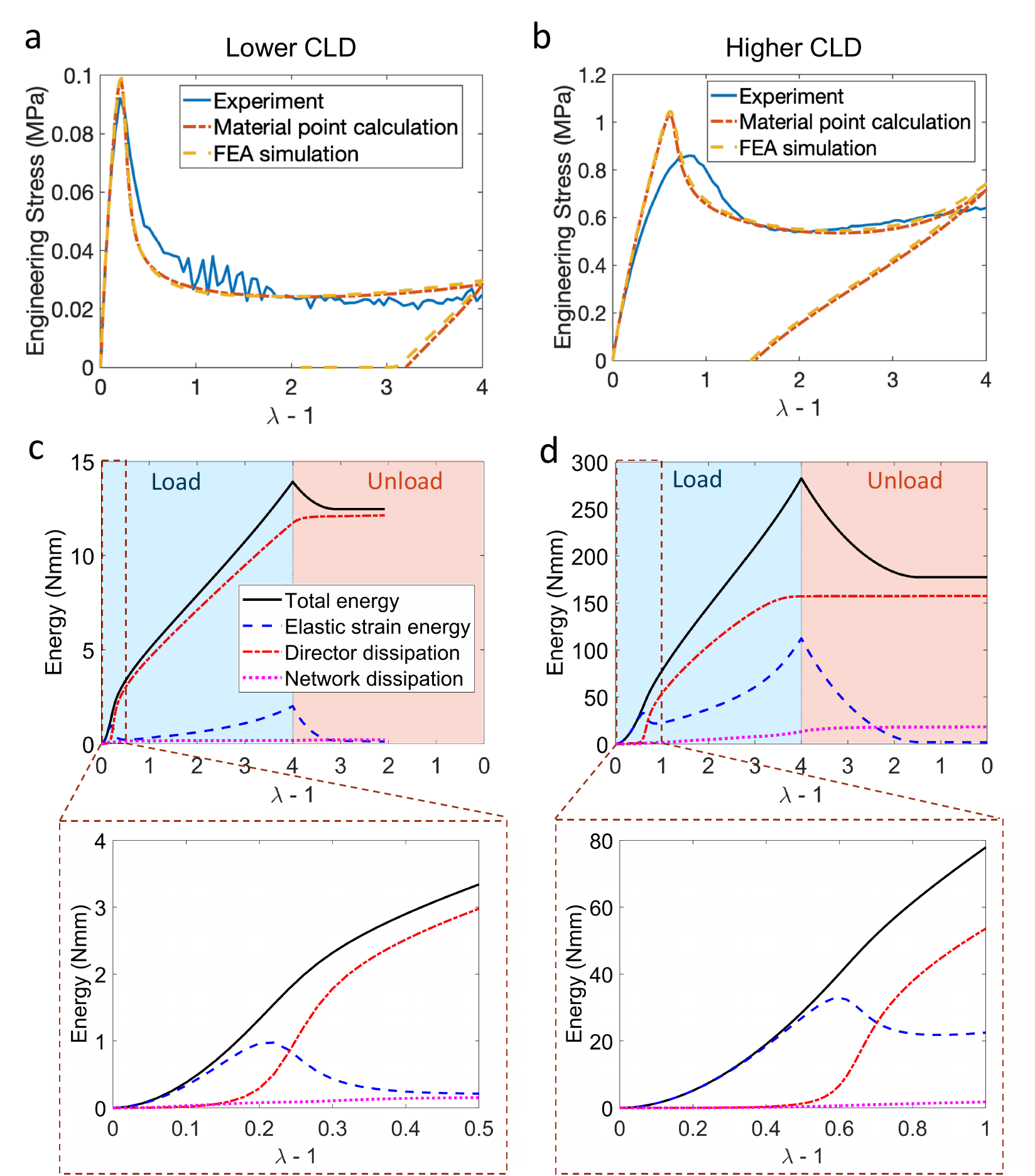}
    \caption{Uniaxial tensile load–unload response at 1\%/s for perpendicular LCEs with (a) lower and (b) higher crosslinking density (CLD), comparing experiment, material point calculation, and FEA simulation. (c, d) Corresponding energy contributions from FEA, showing elastic strain energy, network dissipation, and dissipation from director rotation. The shaded regions indicate loading (blue) and unloading (red). Insets in (c) and (d) show zoomed-in views of the early loading regime to highlight energy accumulation at small strains. Here, \(\lambda\) is the stretch ratio.}
    \label{fig:Fig_5}
\end{figure}

We performed quasi-static uniaxial tension tests to investigate the finite-strain viscoelastic behavior of LCEs with varying crosslinking densities. Figure~\ref{fig:Fig_5}a-b compares engineering stress–strain responses from experiments, material point calculations, and FEA simulations for monodomain LCEs with initial director orientation perpendicular to the loading direction.

The less crosslinked LCEs exhibited a sharp stress drop after the peak, corresponding to rapid director rotation, whereas the more crosslinked samples showed smoother softening. Additional experimental results at other strain rates and for parallel specimens are presented in Figure~\ref{fig:Fig_S2} and discussed in Appendix~\ref{app:UniaxialTension_Rates}. All LCEs exhibited pronounced rate dependence, with nonzero residual strains and significant hysteresis upon unloading, confirming their potential for energy-dissipative applications. DMA measurements (Figures~\ref{fig:Fig_S0}–\ref{fig:Fig_S1}, Appendix~\ref{app:DMA}) revealed $\tan\delta > 0.5$ over a broad frequency range ($10^1$–$10^7$~rad/s), indicating strong viscoelastic damping.

Simulations captured key features of the experimental response, including peak stress and the soft stress plateau. The material point and FEA results closely agreed with each other. The parameters were fitted using experimental data up to 400\% strain, which was sufficient to capture the plateau relevant to the horizontal LCE member deformation in the unit cell (Section~\ref{Sec:UnitCells}). The model parameters are listed in Tables~\ref{tab:Model_SoftPerpLCE}-\ref{tab:Model_StiffPerpLCE}. The fits were performed at a strain rate of 1\%/s, and the model provided reasonable predictions across other rates (Figures~\ref{fig:Fig_S3}-\ref{fig:Fig_S4}).

The elastic strain energy and dissipation calculated for the FEA simulations are shown in Figure~\ref{fig:Fig_5}c-d. For both LCEs, elastic strain energy increased with applied stretch in the early stage. Dissipation from mesogen rotation remained negligible until soft stress activation, after which it rose sharply and eventually surpassed elastic strain energy and network dissipation. Upon unloading, elastic energy decreased, reflecting partial recovery, while dissipation from mesogen rotation remained constant, indicating irreversible rotation. Dissipation from the network viscoelasticity continued to increase during unloading due to time-dependent stress relaxation. The more crosslinked LCE exhibited greater elastic energy storage and higher network dissipation. However, in both cases, dissipation from mesogen rotation was the dominant contributor to total energy loss. Increasing crosslinking density enhanced the stiffness of the network and its relative contribution to energy absorption.

\subsection{Energy absorption of unit cell structures}
\label{Sec:UnitCells}

\begin{figure}[H]
    \centering     
    \includegraphics[width=\textwidth]{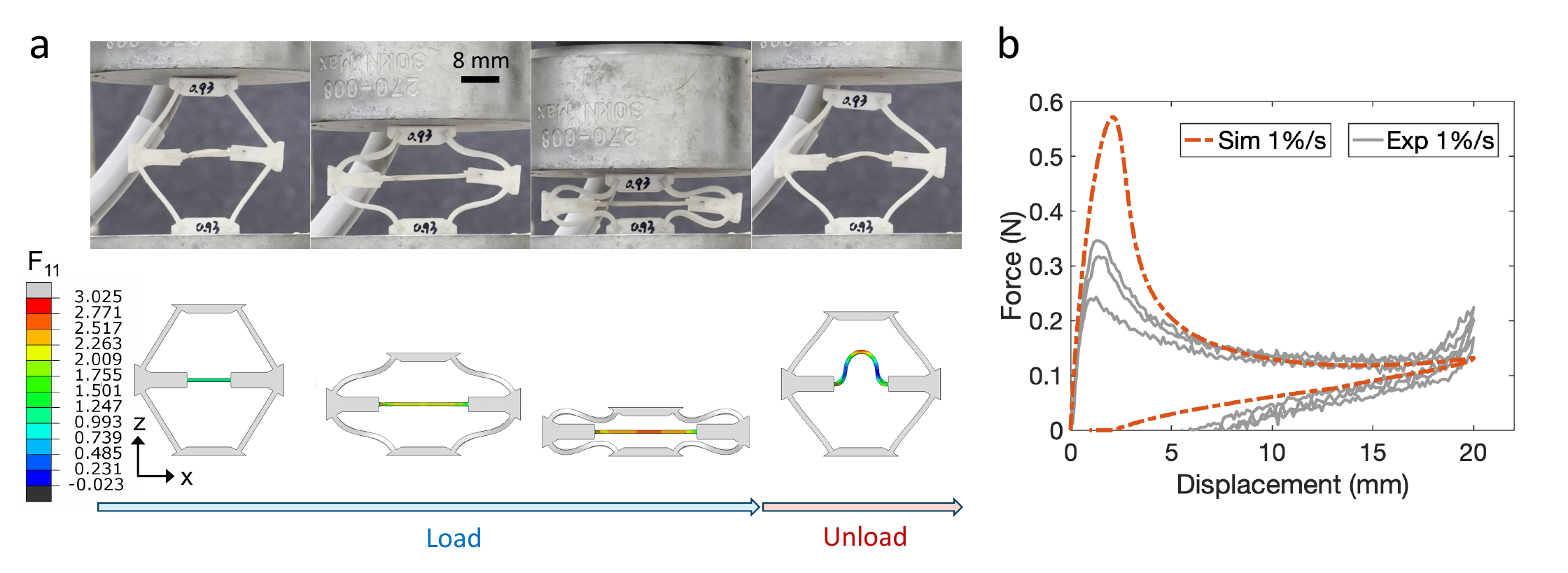}
    \caption{(a) Deformation snapshots from compression tests (top, unit cell 1) and finite element simulations (bottom) at a strain rate of 1\%/s. The color map shows the deformation gradient component $F_{11}$ in the horizontal LCE bar. (b) Corresponding force–displacement responses. Three replicate tests were conducted with thermal resetting before each run.}
    \label{fig:Fig_6}
\end{figure}

Figure~\ref{fig:Fig_6}a compares deformation snapshots from compression tests and finite element simulations of unit cell 1 (Table~\ref{tab:SampleKey}) at a strain rate of 1\%/s. The simulation colormap shows the $x$-component of the deformation gradient, $F_{11}$, in the horizontal LCE bar. Corresponding force–displacement curves are shown in Figure~\ref{fig:Fig_6}b. The use of a soft and highly stretchable LCE bar resulted in stretch-dominated deformation. Both the experimental and simulation results captured the elongation of the horizontal member and bending of the tilted LCE beams. The force–displacement response exhibited an initial linear relationship, followed by a peak and subsequent drop into a plateau. %At small compressive displacements, the tilted beams resisted deformation, and the horizontal bar required a threshold force to initiate stretch, resulting in an initial linear regime up to the peak force. 
The decrease in the force can be attributed to the soft stress response of the horizontal LCE member. Upon unloading, the unit cell recovered the original shape, but the horizontal LCE member exhibited residual elongation from mesogen rotation. The elongation can be reversed by heating the unit cell above the nematic-isotropic transition temperature.  

As shown in Figure~\ref{fig:Fig_7}f, the total energy dissipation was dominated by mesogen rotation in the stretched horizontal LCE bar, with additional contributions from viscoelastic bending in the tilted beams. The simulation slightly overestimated the peak force compared to experiments. In the experiments, the horizontal bar exhibited mild buckling during unloading, which was less pronounced than in the simulations. This discrepancy likely arises from differences in boundary conditions: in the experiments, lateral edges were unconstrained, resulting in more compliant deformation, whereas the simulations enforced symmetric displacements and rotational constraints, producing a stiffer response.

\subsubsection{Effect of a stretchable LCE horizontal bar}
\label{Sec:StretchableLCE}

\begin{figure}[H]
    \centering     
    \includegraphics[width=1\textwidth]{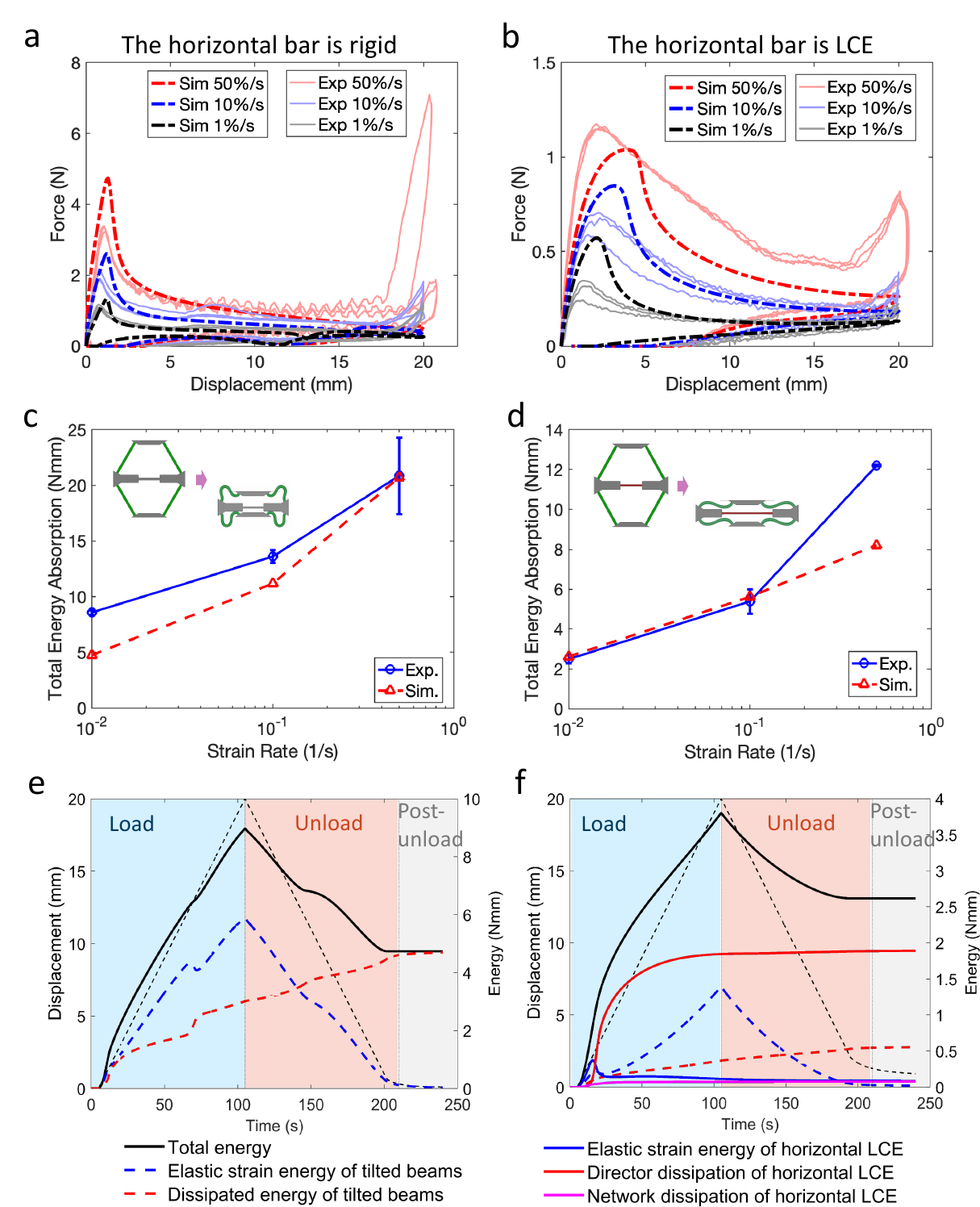}
    \caption{Experimental and simulation results for unit cell structures under compression at strain rates of 1\%/s, 10\%/s, and 50\%/s. (a, c, e) Rigid-bar structure; (b, d, f) stretchable LCE-bar structure. Solid curves represent experiments; dashed curves represent simulations. In (e) and (f), black dashed lines (left axis) show applied displacement, and colored curves (right axis) indicate simulated energy contributions at 1\%/s, including elastic storage, viscoelastic bending, and dissipation from director rotation.}
    \label{fig:Fig_7}
\end{figure}

Figure~\ref{fig:Fig_7} compares the deformation and energy absorption behavior of unit cell structures with either a rigid or stretchable LCE horizontal bar, based on simulations and experiments conducted at strain rates of 1\%/s, 10\%/s, and 50\%/s.

In the rigid-bar structure (Figure~\ref{fig:Fig_7}a, c, e), deformation was governed by buckling of the tilted LCE beams, consistent with prior work~\citep{Jeon2022SynergisticElastomersb}.

Initial loading produced a steep force increase, followed by buckling-induced softening and a secondary force peak near full compression. Energy dissipation primarily arose from viscoelastic bending of the tilted beams, which underwent buckling but ultimately recovered due to viscoelastic relaxation. Simulations captured the overall force–displacement trends, though peak forces were slightly overestimated, particularly at higher strain rates. At 50\%/s, the abrupt increase in the force near the end of loading in experiments was attributed to contact between buckled beams and horizontal supports, enabled by lateral motion and minor geometric asymmetries. Simulated energy absorption showed qualitative agreement with experiments, with viscous dissipation reaching approximately 4.7~N$\cdot$mm at 1\%/s.

In contrast, the compliant LCE horizontal bar in the stretching-dominated structure (Figure~\ref{fig:Fig_7}b, d, f)   elongated significantly, while the tilted beams bent without buckling. This produced a lower peak force followed by a broad plateau. A small rise in the force near the end of loading at 50\%/s was again observed in experiments due to contact with the supports.  % Increased noise at this rate likely arose from inertial effects near the MTS crosshead speed limit and the low reaction force (up to 1~N) relative to the 100~N load cell capacity. 
%Dissipation was dominated by soft stress from director rotation in the horizontal LCE bar, with reduced contributions from beam bending. At 1\%/s, viscous dissipation from beam bending dropped to approximately 0.5~N$\cdot$mm. 
The simulation results for the force-displacement response and the energy absorption of both unit cell structures agreed well with experiments. The simulation results for the energy absorption at the highest rate 50\%/s for the stretchable unit cell underestimated the experimental data.  The model only used a single relaxation process to describe the viscoelastic network behavior of the LCE horizontal member.  The network relaxation time was calibrated using data from a low strain rate test (1\%/s), and thus may not accurately capture the viscoelastic behavior at the highest strain rate of 50\%/s.

Introducing a stretchable LCE bar fundamentally altered the dominant dissipation mechanisms of the unit cell. While mesogen rotation enabled a new dissipation mechanism, lateral expansion suppressed beam buckling and reduced viscoelastic bending. At 1\%/s, the total energy absorption decreased from 4.7~N$\cdot$mm (rigid-bar) to 2.5~N$\cdot$mm (stretchable-bar). This highlights a key trade-off: stiff horizontal members promote buckling-driven energy absorption, whereas compliant ones enable tension-driven dissipation from mesogen rotation. Optimizing the balance between these mechanisms is essential for maximizing energy absorption and is explored in the next section.

\subsubsection{Effect of horizontal bar thickness and material properties}
\label{sec:ThicknessRatio}

As shown in Section~\ref{Sec:StretchableLCE}, introducing a stretchable LCE horizontal bar adds energy dissipation via mesogen rotation but suppresses bending in the tilted beams, a competing dissipation mechanism. To explore this trade-off, we conducted a parametric study in which we varied the horizontal bar thickness \(t_h\) while holding all other geometric and material parameters constant.

\begin{figure}[H]
    \centering     
    \includegraphics[width=0.98\textwidth]{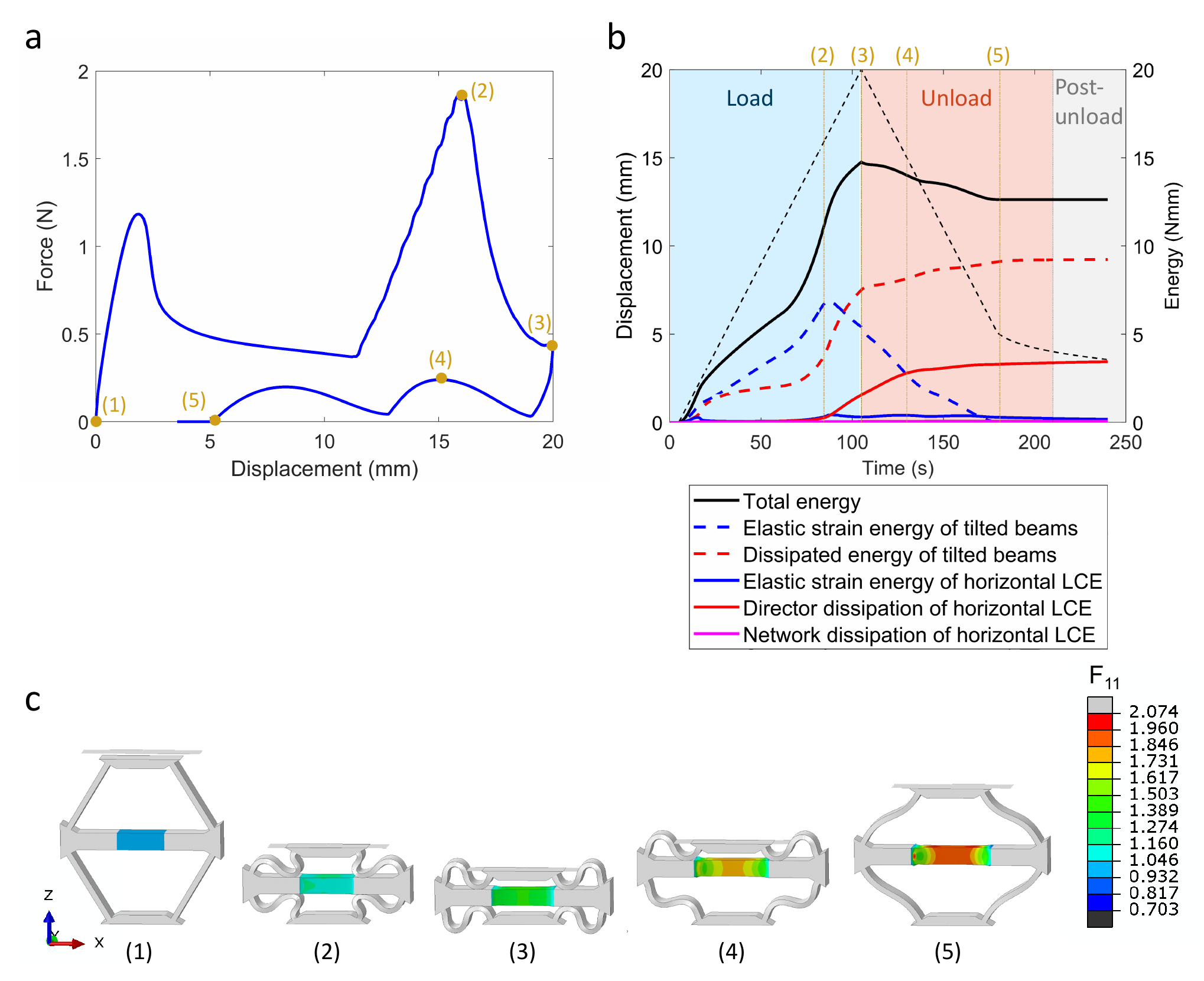}
    \caption{Finite element simulation of a unit cell structure with a 3~mm thick horizontal LCE bar (\(t_h/t_t = 3.23\)) under compression at 1\%/s. (a) Force–displacement curve. (b) Evolution of energy contributions from elastic storage, viscoelastic beam bending, and director rotation. The black dashed line indicates the applied displacement. (c) Deformation snapshots colored by the $x$-component of the deformation gradient ($F_{11}$) in the horizontal LCE bar. Sequential activation of bending and soft stress enabled total energy absorption exceeding that of the rigid-bar case.}
    \label{fig:Fig_8}
\end{figure}

Figure~\ref{fig:Fig_8} shows the simulation results for the case with \(t_h = 3\)~mm (\(t_h/t_t = 3.23\)) under compression at 1\%/s. At $t_h=3$ mm, the horizontal bar was stiff enough that deformation was initially dominated by the buckling of the tilted beams. 
The primary contributions to the energy absorption initially are the elastic energy and viscous dissipation solely from the tilted beams (Figure~\ref{fig:Fig_8}b). After the beams have buckled, further compression triggered significant stretching of the horizontal LCE bar, activating director rotation and the soft stress response. Notably, the system is designed such that the tension-induced dissipation in the horizontal member is engaged after the beams have buckled. From this point to the end of loading (stage (2) to (3)), both soft stress and bending-related dissipation increased concurrently.
During unloading, the two layers of buckled beams recovered sequentially, causing the adjacent layers to exhibit a cyclic recovery and reloading behavior, consistent with our prior study ~\citep{Jeon2022SynergisticElastomersb}. This reduced the elastic strain energy and increased viscous network dissipation contributions to the total energy. Simultaneously, the horizontal bar continued to stretch, adding further dissipation via director rotation. The combined effect yielded a total energy absorption of  13~N$\cdot$mm, which was over 2.5 times that of the rigid-bar case, demonstrating the benefit of coupling both mechanisms.

\begin{figure}[H]
    \centering
    \includegraphics[width=0.885\textwidth]{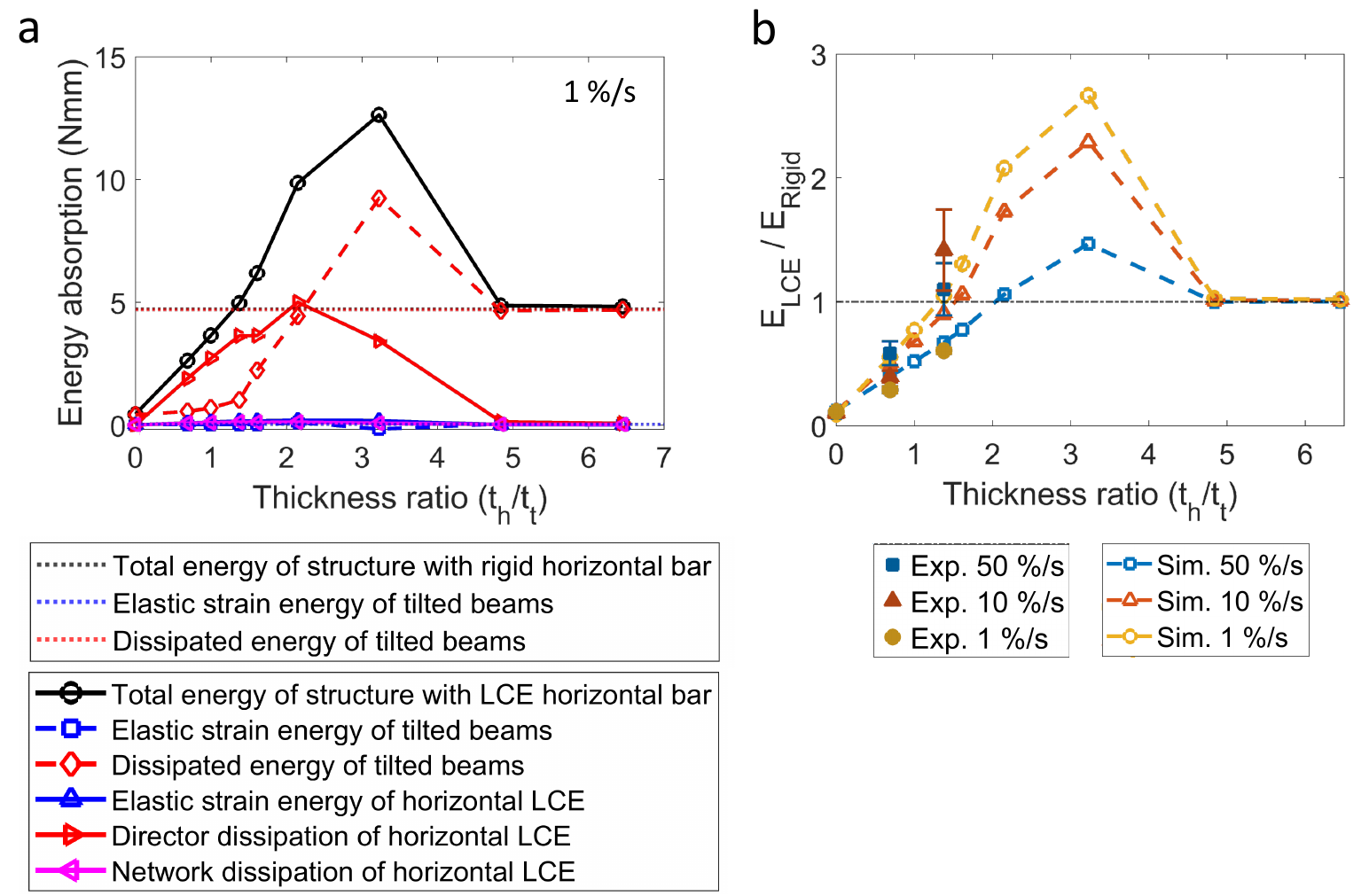}
    \caption{Effect of thickness ratio \(t_h/t_t\) between the horizontal and tilted beams on energy absorption. (a) Energy contributions for structures with a rigid bar (thin dotted lines) and stretchable LCE bar (thick lines), at 1\%/s. The LCE-bar structure shows a non-monotonic dependence, peaking at intermediate \(t_h/t_t\). (b) Normalized total energy absorption \(E_\mathrm{LCE}/E_\mathrm{Rigid}\) at 1\%/s, 10\%/s, and 50\%/s. Open symbols: simulations; filled symbols: experiments. The optimal thickness ratio decreases at lower strain rates.}
    \label{fig:Fig_9}
\end{figure}

We repeated the simulation for varying ratios of the thickness of the horizontal bar and tilted beams \(t_h/t_t\). As shown in Figure~\ref{fig:Fig_9}a, the energy absorption of the LCE-bar structure first increased with the thickness ratio. The energy absorption reached a maximum at \(t_h/t_t = 3\), then decreased to equal the energy absorption of the rigid horizontal bar unit cell for \(t_h/t_t > 3\).
For small \(t_h/t_t\), the LCE horizontal bar was too compliant, resulting in a stretch-dominated response with minimal bending of the tilted beams and, consequently, low energy absorption. In this regime, dissipation was dominated by director rotation in the horizontal bar. This is because the beams primarily rotated with little bending to accommodate the large stretch of the thin horizontal bar. As \(t_h/t_t\) increased, the horizontal bar became stiffer, promoting greater bending, then buckling of the tilted beams prior to the onset of director rotation and the associated soft stress response in the horizontal bar. This transition is reflected in the rising energy dissipation from both the tilted beams and the director rotation of the horizontal bar. This sequential activation of viscoelastic buckling and director rotation enabled the overall energy absorption of the stretchable horizontal-bar unit cell to exceed that of the rigid-bar structure by nearly a factor of three. Compared to smaller \(t_h/t_t\), the dissipation of tilted beams contributed significantly more due to increased bending, while horizontal bar stretch remained sufficient to sustain high levels of dissipation from director rotation. The director rotation–induced dissipation peaked around \(t_h/t_t = 2\). Beyond this point, further increases in bar stiffness enhanced beam buckling but reduced stretch in the horizontal bar, leading to a decline in director rotation and associated dissipation. As the horizontal bar became excessively stiff, it could no longer stretch enough to activate director rotation, causing this dissipation mode to diminish. Meanwhile, dissipation from the tilted beams gradually converged to that of the rigid-bar case. Note that network dissipation in the LCE horizontal bar was negligible at these strain levels, as director rotation dominated the response. However, if the bar were stretched beyond the soft stress regime, network dissipation would become more significant.

The normalized energy ratio \(E_\text{LCE}/E_\text{Rigid}\), plotted in Figure~\ref{fig:Fig_9}b, shows that LCE-bar structures outperformed the rigid-bar baseline once \(t_h/t_t\) exceeded a threshold. This optimal ratio shifted to smaller values at lower strain rates, highlighting the influence of viscoelasticity. At higher strain rates, the benefit of having a stretchable horizontal bar diminishes; the bar becomes effectively too stiff to stretch, and energy absorption approaches that of the rigid-bar case. Experimental results from unit cells 1, 3, and 5 qualitatively followed simulation trends. Deviations at larger \(t_h/t_t\) ratios likely resulted from boundary condition differences: lateral rotation and asymmetry were observed in experiments but constrained in simulations.

\begin{figure}[H]
    \centering     
    \includegraphics[width=0.885\textwidth]{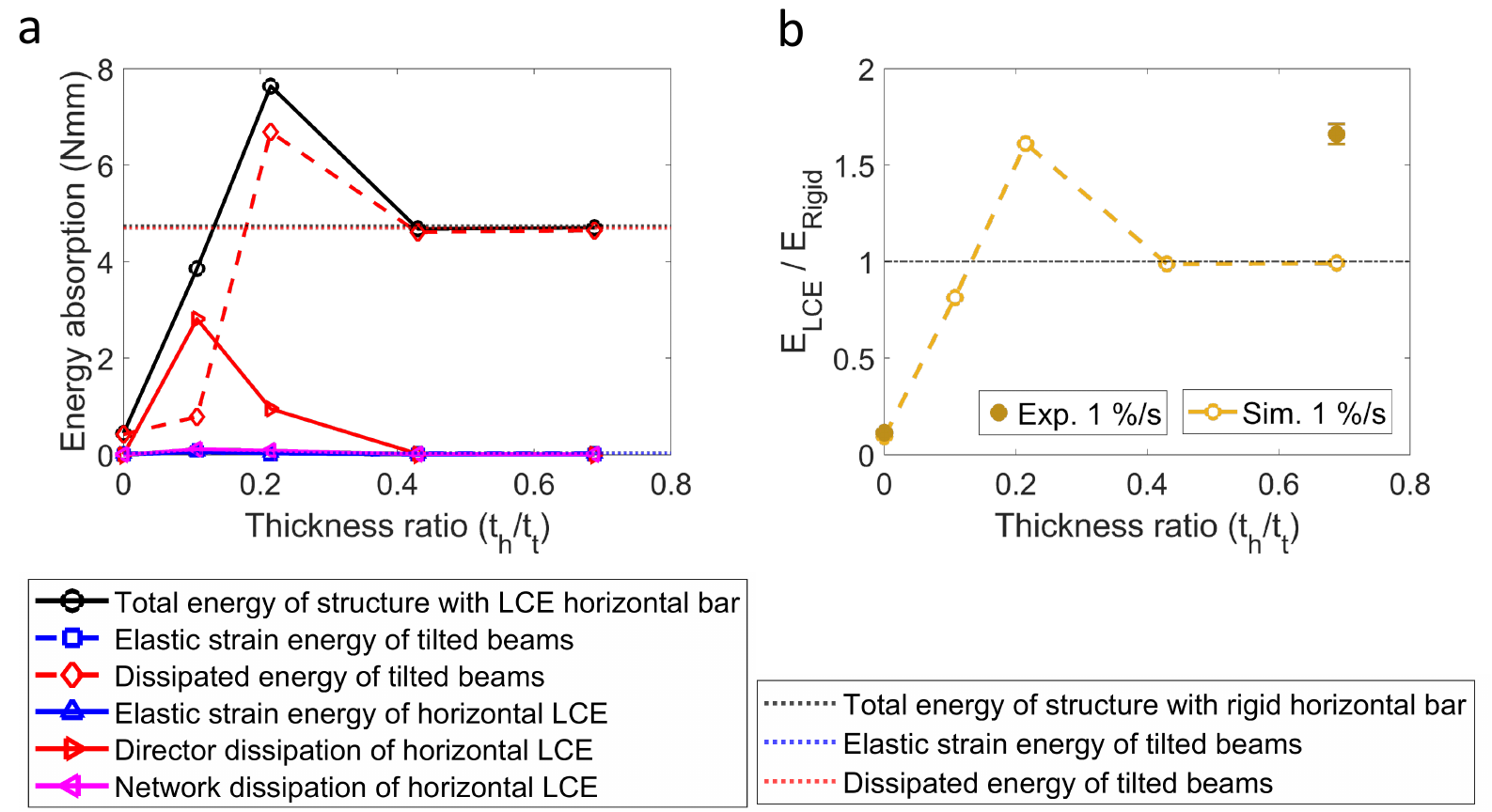}
    \caption{Effect of horizontal LCE bar material. (a) Energy absorption versus thickness ratio \(t_h/t_t\) for structures with either a rigid bar (thin dotted lines) or a stiffer perpendicular LCE bar (thick lines), at 1\%/s. (b) Energy absorption ratio \(E_\mathrm{LCE}/E_\mathrm{Rigid}\). Open circles: simulations; filled circles: experiments. Simulations show a non-monotonic trend with a narrower optimum for the stiffer LCE horizontal bar.}
    \label{fig:Fig_10}
\end{figure}

To evaluate the effect of material properties in the tilted beams and horizontal bar, we repeated the analysis using a stiffer horizontal bar composed of perpendicular LCE with higher crosslinking density (parameters listed in Table~\ref{tab:Model_StiffPerpLCE}). As shown in Figure~\ref{fig:Fig_10}, energy absorption retained a non-monotonic dependence on \(t_h/t_t\), but the optimal ratio shifted to a lower value and the effective range narrowed due to the increased stiffness. Two experimental points are included for comparison. At \(t_h/t_t = 0\), simulations and experiments showed good agreement. At higher \(t_h/t_t\), simulations predicted symmetric buckling, whereas experiments exhibited rotation and asymmetric buckling of the tilted beams. This discrepancy is mainly attributed to boundary conditions: in simulations, the lateral edges of the unit cell were constrained against rotation, whereas in experiments, they were free to deform.

Overall, the parametric studies on horizontal bar thickness and material properties underscore the importance of tuning both geometry and material selection to balance viscoelastic bending and soft-stress tensile response for enhanced energy absorption.
 %When optimized, architected LCE structures can surpass conventional snap-through designs and achieve energy dissipation beyond what either mechanism offers alone.

\begin{figure}[H]
    \centering     
    \includegraphics[width=0.97\textwidth]{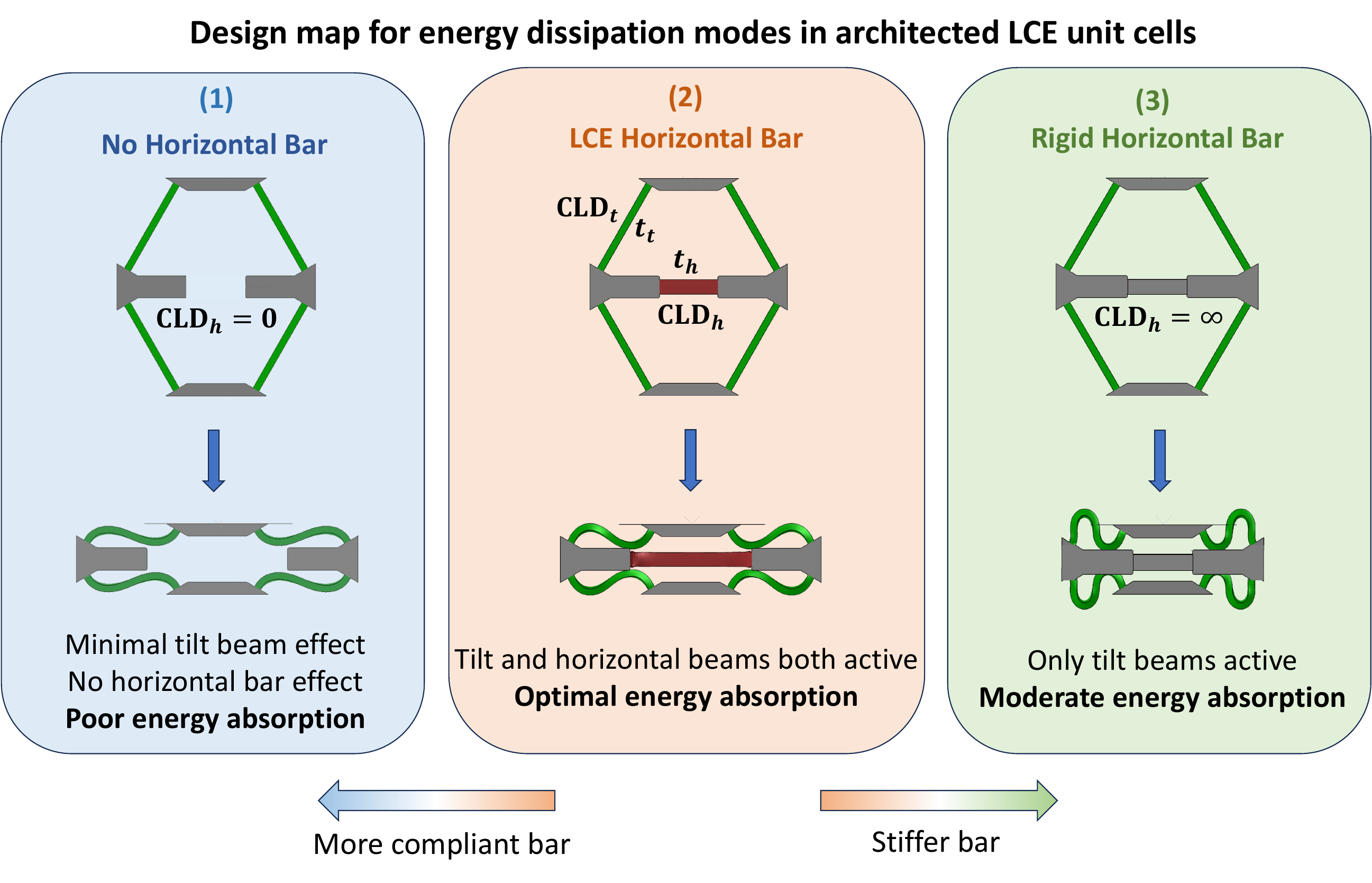}
    \caption{Design map illustrating energy dissipation regimes in architected LCE unit cells with varying horizontal bar properties. Configuration (1) lacks a horizontal bar and exhibits low energy absorption. Configuration (3) uses a rigid horizontal bar, leading to moderate energy absorption dominated by tilted beam bending. Configuration (2) achieves maximal dissipation by balancing viscoelastic bending in tilted beams with tension-induced soft stress in a stretchable horizontal LCE bar. Transitions between regimes are governed by the in-plane thickness ratio \( t_h/t_t \), applied strain rate, and the crosslinking density ratio \( \mathrm{CLD}_h/\mathrm{CLD}_t \).}
    \label{fig:Fig_11}
\end{figure}

The key trends are synthesized schematically in Figure~\ref{fig:Fig_11}, which illustrates three configurations and their associated energy dissipation mechanisms. Configuration (1) lacks a horizontal bar and exhibits poor energy absorption due to negligible tensile resistance and minimal beam bending. Configuration (3) employs a rigid horizontal bar that suppresses axial stretch, limiting dissipation to viscoelastic buckling of the tilted beams. In contrast, Configuration (2) achieves optimal energy absorption by coupling tension-induced soft stress in a stretchable LCE horizontal bar with viscoelastic bending in the tilted beams. This schematic emphasizes the need to tailor the structural stiffness of the tension element relative to the flexural stiffness of the bending element, for example by modifying the thickness, elastic moduli, and strain rate,  to effectively harness both dissipation mechanisms.

\section{Conclusion}

This study presents a novel strategy for enhancing energy dissipation in architected materials by leveraging the soft stress response of liquid crystal elastomers (LCEs). Specifically, we introduced stretchable LCE horizontal bars into hexagonal lattices with tilted LCE beams, enabling tension-induced mesogen rotation to contribute to dissipation under macroscopic compression.

We first characterized the finite-strain viscoelastic behavior of LCEs with varying crosslinking densities through dynamic mechanical analysis (DMA), uniaxial tension tests, and finite element simulations. Simulations employed a custom Abaqus UEL based on nonlinear viscoelastic theory for nematic LCEs, capturing the rate-dependent deformation and soft stress response. Dissipation in monodomain LCEs stretched into the soft stress plateau was dominated by viscous director rotation, while higher crosslinking densities led to stiffer responses, greater elastic energy storage, and enhanced network dissipation.

Building on prior work with rigid horizontal bars~\citep{Jeon2022SynergisticElastomersb}, we introduced compliant LCE bars to activate dissipation via mesogen rotation under tensile deformation. However, this mechanism reduced beam bending by decreasing the tilt angle during compression, revealing a competition between the two modes. Through experiments and simulations, we demonstrated a non-monotonic dependence of energy absorption on the thickness ratio between horizontal and tilted beams. Thin bars promoted stretch-dominated deformation, whereas thick bars behaved as rigid constraints. At an intermediate, optimal ratio, sequential activation of viscoelastic beam bending followed by bar stretching yielded energy absorption 2–3 times greater than the rigid-bar case, exceeding the additive contributions of each mechanism. We further showed that increasing the crosslinking density in the horizontal bar shifted the optimal thickness ratio lower, and that lower strain rates enhanced energy dissipation and broadened the optimal design window.

These findings establish a design framework for coupling viscoelastic buckling with tension-driven soft stress in LCE-based lattices. By tuning geometric and material parameters, energy absorption can surpass that of conventional buckling instability-driven designs, with implications for impact mitigation, damping, and protective systems.

Several limitations remain. Simulated boundary conditions constrained lateral rotation, unlike the free deformation observed experimentally. Even in optimized cases, horizontal bar stretch remained limited, preventing full utilization of the soft stress regime. Parameters such as the stretchable bar length and beam tilt angle were not systematically explored. Snap-through buckling, which could trap elastic energy, was not investigated. Network dissipation in the horizontal bar was minimal at the applied strain levels but may be enhanced by designs that stretch beyond the soft stress regime.
Nonetheless, to our knowledge, this is the first in-depth study to combine tension-induced mesogen rotation of LCEs with viscoelastic beam buckling in structural-level energy absorption. By revealing the interplay and competition between these mechanisms, this work lays a foundation for future exploration of LCE-based architected materials that extends beyond conventional design paradigms for energy absorption.

\section*{CRediT authorship contribution statement}
\textbf{Beijun Shen:} Conceptualization, Methodology, Investigation, Formal analysis, Software, Validation, Data Curation, Visualization, Writing - Original Draft, Writing - Review \& Editing.
\textbf{Yuefeng Jiang:} Software, Validation, Writing - Review \& Editing.
\textbf{Christopher M. Yakacki:} Conceptualization, Methodology, Writing - Review \& Editing.
\textbf{Sung Hoon Kang:} Conceptualization, Supervision, Funding acquisition, Writing - Review \& Editing.
\textbf{Thao D. Nguyen:} Conceptualization, Methodology, Resources, Supervision, Project administration, Funding acquisition, Writing - Review \& Editing.

\section*{Declaration of competing interest}
The authors declare that they have no competing financial interests or personal relationships that could have appeared to influence the work reported in this paper.

\section*{Data availability}
Data will be made available on request.

\section*{Acknowledgements}
The authors would like to thank Prof. Sanjay Govindjee from the Department of Civil and Environmental Engineering at the University of California, Berkeley, for his insightful discussions and suggestions. B.S. gratefully acknowledges Dr. Alex Sun for his thoughtful guidance and unwavering support during challenging times. B.S. also thanks Dr. Tingting Xu and Mr. Bibekananda Datta for their valuable discussions and feedback. 
This work was supported in part by the Army Research Office (Grant Number W911NF-17-1-0165) (C.M.Y., S.H.K., T.D.N), Hanwha Non-tenured Faculty Award (S.H.K.), KAIST Start-Up Fund (S.H.K.), and Brain Pool Plus program through the National Research Foundation of Korea (NRF) funded by the Ministry of Science, ICT and Future Planning (award number: RS-2024-00439827) (S.H.K.).


\begin{appendices}

\renewcommand{\thetable}{\Alph{section}\arabic{table}}
\renewcommand{\thefigure}{\Alph{section}\arabic{figure}} 
\numberwithin{equation}{section}

\section{Characterization of viscoelastic properties of LCEs}
\label{app:DMA}
\setcounter{table}{0}
\setcounter{figure}{0}

\begin{figure}[H]
    \centering
    \includegraphics[width=1\textwidth]{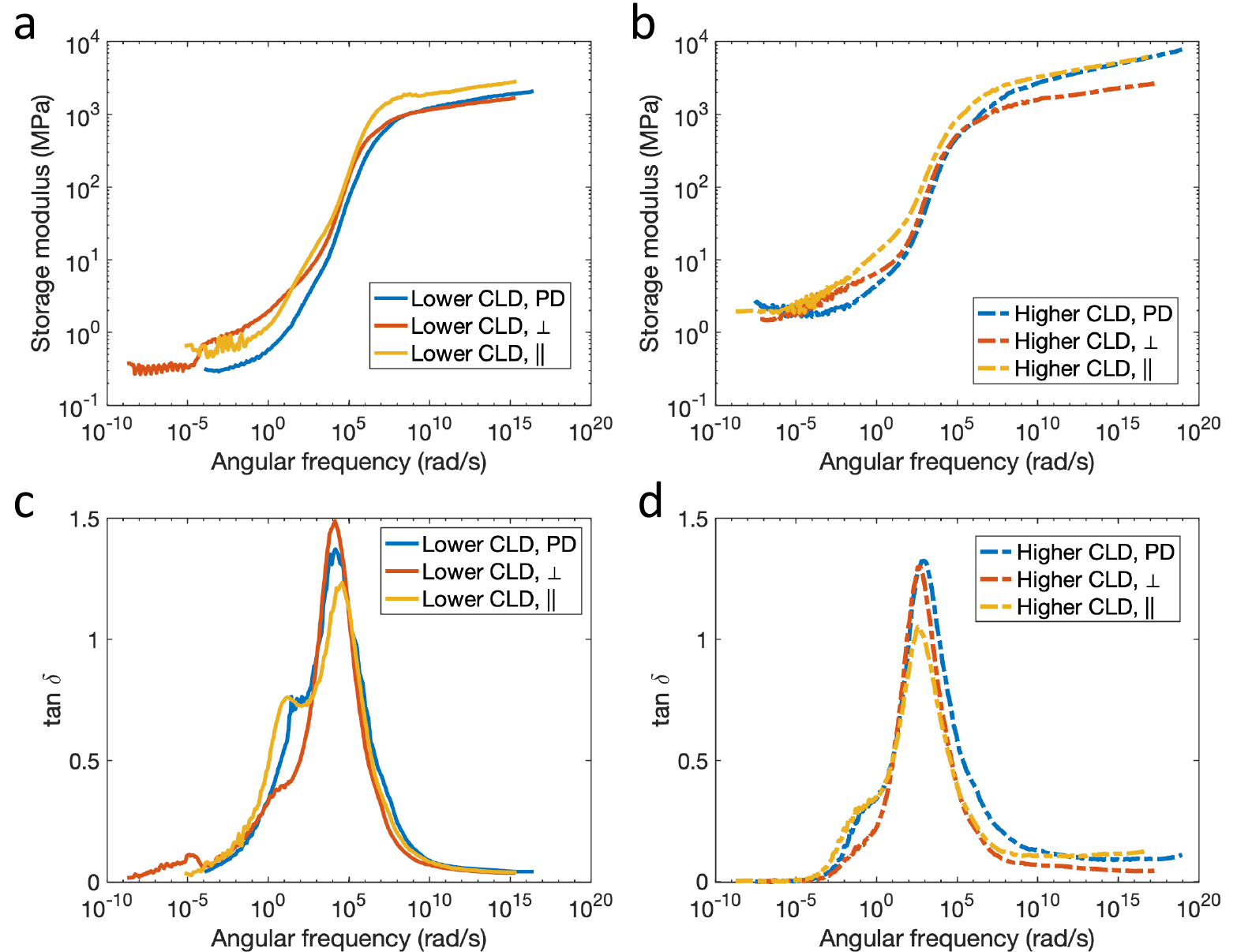}
    \caption{DMA characterization of polydomain (PD), perpendicular ($\perp$), and parallel ($\parallel$) LCEs. In $\perp$ and $\parallel$ samples, the initial director was aligned perpendicular or parallel to the loading direction, respectively. (a, b) Master curves of storage modulus for lower and higher crosslinking density. (c, d) Corresponding loss factor ($\tan\delta$) curves. All data referenced to 25~$^\circ$C.}
    \label{fig:Fig_S0}
\end{figure}

Figure~\ref{fig:Fig_S0} shows the master curves of storage modulus and loss factor ($\tan\delta$) for LCEs with varying crosslinking densities and director alignments. The storage modulus spanned over five orders of magnitude—from approximately $10^{-1}$ to $10^4$~MPa—across frequencies ranging from $10^{-5}$ to $10^{15}$~rad/s. LCEs with higher crosslinking density exhibited consistently higher moduli. Among different alignments, polydomain LCEs showed the lowest modulus at low frequencies under matched crosslinking conditions.

\begin{figure}[H]
    \centering
    \includegraphics[width=1\textwidth]{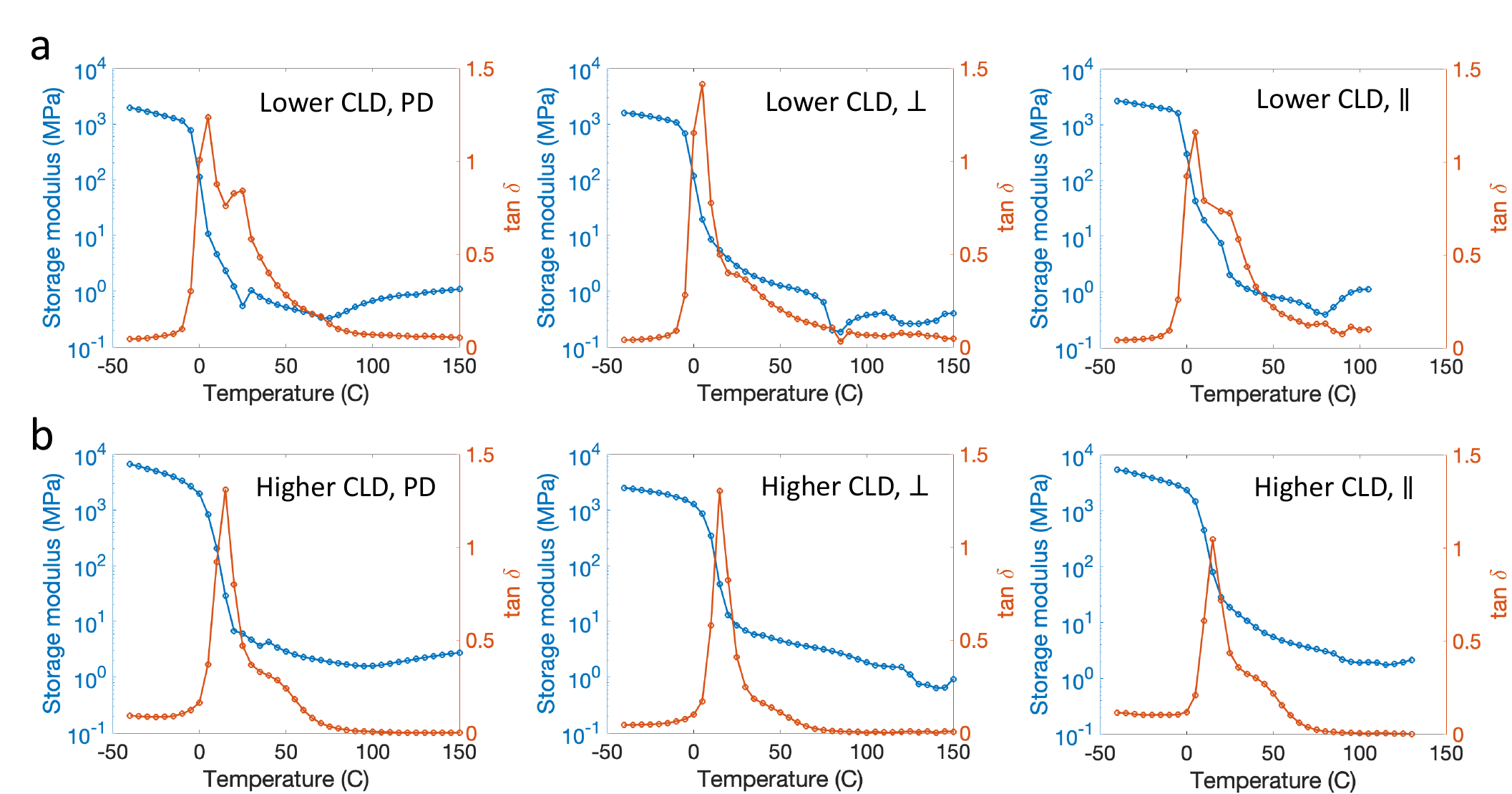}
    \caption{Storage modulus and $\tan\delta$ vs.\ temperature at 0.93~Hz for LCEs with (a) lower and (b) higher crosslinking density (CLD).}
    \label{fig:Fig_S1}
\end{figure}

All samples exhibited broad dissipation behavior, with $\tan\delta$ exceeding 0.5 between $10^1$ and $10^7$~rad/s. LCEs with higher crosslinking density had reduced peak $\tan\delta$ values and a shift to lower frequencies. A secondary peak with an extended shoulder was present across all samples, though it was smaller for more crosslinked networks. Parallel LCEs displayed lower primary $\tan\delta$ peaks compared to polydomain and perpendicular counterparts, whereas their shoulder heights remained comparable to those of polydomain samples. Perpendicular samples showed the lowest shoulders. Figure~\ref{fig:Fig_S1} presents the corresponding temperature-dependent results. All LCEs showed characteristic transitions: a modulus drop near the glass transition temperature ($T_\text{g}$), a broad dissipation peak between $T_\text{g}$ and the nematic-to-isotropic transition temperature ($T_\text{ni}$), and a secondary modulus drop near $T_\text{ni}$ due to dynamic soft elasticity~\citep{Yakacki2015TailorableReaction, Saed2016SynthesisReaction}. Increasing crosslinking density raised both $T_\text{g}$ and $T_\text{ni}$—from approximately 5~$^\circ$C to 15~$^\circ$C for $T_\text{g}$, and from 80~$^\circ$C to 100~$^\circ$C for $T_\text{ni}$—and increased moduli across all temperatures. These trends align with prior literature~\citep{Saed2017Thiol-acrylateStrain, Merkel2018ThermomechanicalElastomers, Traugutt2017Liquid-crystalBehavior, Saed2016SynthesisReaction}.

\section{Finite-strain uniaxial tension load-unload tests of LCEs}
\label{app:UniaxialTension_Rates}
\setcounter{table}{0}
\setcounter{figure}{0}

\begin{figure}[H]
    \centering
    \includegraphics[width=0.9\textwidth]{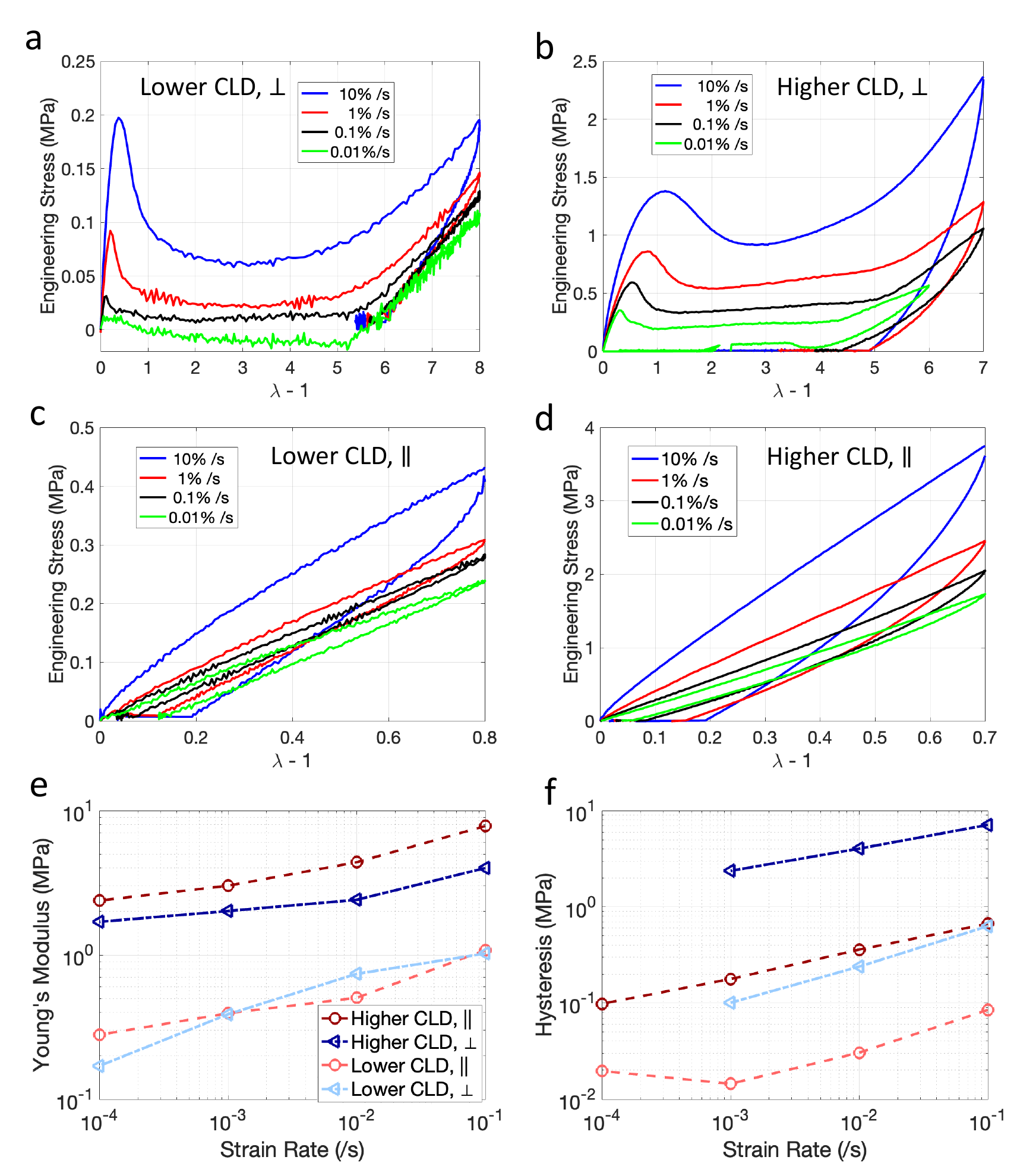}
    \caption{Uniaxial load–unload tests at varying strain rates for LCEs with different crosslinking densities (CLD). Engineering stress–strain responses for (a–b) perpendicular and (c–d) parallel LCEs. In the perpendicular and parallel configurations, the director is initially perpendicular or aligned with the loading direction. (e) Young’s modulus and (f) hysteresis as functions of strain rate. Stretch is denoted by $\lambda$.}
    \label{fig:Fig_S2}
\end{figure}

Figure~\ref{fig:Fig_S2} shows the rate-dependent uniaxial stress–strain responses of perpendicular (a–b) and parallel (c–d) LCEs with different crosslinking densities (CLD). Variability across repeats was minimal; all specimens were thermally reset before testing. All LCEs exhibited pronounced strain rate dependence and retained residual strain after unloading. Recovery at room temperature occurred within 1–2 minutes for more crosslinked samples and 5–10 minutes for less crosslinked ones. Full recovery occurred immediately above $T_\text{ni}$.

Perpendicular LCEs showed an initial linear regime up to peak stress, followed by softening from mesogen rotation and stiffening from chain alignment. Less crosslinked samples exhibited a sharp stress drop after the peak, whereas more crosslinked ones showed smoother transitions. At the lowest rate (0.01\%/s), forces in the low-CLD sample fell below the reliable detection limit of the 500~N load cell. Both peak and flow stresses increased with strain rate, consistent with stress-activated director rotation theory~\citep{Wang2022AElastomers}. The plateau widened at slower rates, and residual strain matched the plateau strain. Hysteresis increased with rate, especially in the stiffening region, indicating substantial polymer network viscoelasticity, supported by broad $\tan\delta$ plateaus in Figures~\ref{fig:Fig_S0}-\ref{fig:Fig_S1}.
Parallel samples (Figure~\ref{fig:Fig_S2}c–d) also showed rate-dependent behavior but lacked soft stress, as the directors were aligned with the loading axis. Dissipation was governed by network viscoelasticity, as indicated by load–unload hysteresis.

Figures~\ref{fig:Fig_S2}e–f quantify Young’s modulus and hysteresis as functions of strain rate. Both followed power-law scaling. Parallel LCEs were generally stiffer, especially in the high-CLD network where $E_{\parallel} \approx 1.5$–$2E_{\perp}$, consistent with prior reports~\citep{MartinLinares2020TheElastomer, Merkel2018ThermomechanicalElastomers}. In contrast, low-CLD LCEs showed modest anisotropy.
At 1\%/s, some modulus estimates in the low-CLD group were less reliable due to near-threshold forces. Storage modulus trends from DMA (Figure~\ref{fig:Fig_S0}) further support these anisotropic mechanical responses.

Hysteresis increased with strain rate across all conditions (Figure~\ref{fig:Fig_S2}f), but comparison between perpendicular and parallel samples is complicated by different maximum strains. Perpendicular LCEs were extended into the mesogen rotation and strain-stiffening regimes, resulting in larger hysteresis. An outlier in the low-CLD parallel group at 0.01\%/s may reflect noise at low force levels.

\section{Material point modeling and parameter fitting for perpendicular LCEs during uniaxial tension}
\label{app:MaterialPoint}
\setcounter{table}{0}
\setcounter{figure}{0}

To model the finite-strain, rate-dependent behavior of perpendicular LCEs, we employed the nonlinear viscoelastic framework developed by Wang et al.~\citep{Wang2022AElastomers}, which captures the coupling between director rotation and polymer network viscoelasticity. Material point simulations assumed homogeneous uniaxial deformation, enabling efficient parameter identification while retaining key physical mechanisms. Simulations were conducted at strain rates of 10\%/s, 1\%/s, 0.1\%/s, and 0.01\%/s for two LCEs with different crosslinking densities. Model predictions were compared to experimental results (Figures~\ref{fig:Fig_S2}a-b).

\begin{table}[H]
\centering
\caption{Model parameters for perpendicular LCE with lower crosslinking density (CLD).}
\small
\begin{tabular}{@{}ccc@{}}
    \toprule
    \textbf{Parameter} & \textbf{Physical significance} & \textbf{Value} \\ 
    \midrule
    $Q$ & Order parameter & 0.87  \\ 
    $\mu^\mathrm{eq}$ & Equilibrium shear modulus & 0.20 MPa \\ 
    $\mu^\mathrm{neq}$ & Nonequilibrium shear modulus & 0.05 MPa \\ 
    $\eta_N$ & Network viscosity & 0.5 MPa$\cdot$s \\ 
    $I_m$ & Stiffening parameter & 1000 \\ 
    $\eta_{D_0}$ & Initial director viscosity & 11 MPa$\cdot$s \\ 
    $k_S$ & Viscosity evolution coefficient & 0.1 \\ 
    $\theta_0$ & Initial director angle & 83$^\circ$ \\ 
    $\kappa$ & Bulk modulus & 4 MPa \\ 
    $\gamma$ & Penalty enforcing $||\bm{d}||=1$ & 2 \\
    \bottomrule
\end{tabular}
\label{tab:Model_SoftPerpLCE}
\end{table}

\begin{table}[H]
\centering
\caption{Model parameters for perpendicular LCE with higher crosslinking density (CLD).}
\small
\begin{tabular}{@{}ccc@{}}
    \toprule
    \textbf{Parameter} & \textbf{Physical significance} & \textbf{Value} \\ 
    \midrule
    $Q$ & Order parameter & 0.62  \\ 
    $\mu^\mathrm{eq}$ & Equilibrium shear modulus & 0.55 MPa \\ 
    $\mu^\mathrm{neq}$ & Nonequilibrium shear modulus & 0.25 MPa \\ 
    $\eta_N$ & Network viscosity & 700 MPa$\cdot$s \\ 
    $I_m$ & Stiffening parameter & 10 \\ 
    $\eta_{D_0}$ & Initial director viscosity & 1000 MPa$\cdot$s \\ 
    $k_S$ & Viscosity evolution coefficient & 0.6 \\ 
    $\theta_0$ & Initial director angle & 80$^\circ$ \\ 
    $\kappa$ & Bulk modulus & 11 MPa \\ 
    $\gamma$ & Penalty enforcing $||\bm{d}||=1$ & 2 \\
    \bottomrule
\end{tabular}
\label{tab:Model_StiffPerpLCE}
\end{table}

\begin{figure}[H]
    \centering
    \includegraphics[width=0.77\textwidth]{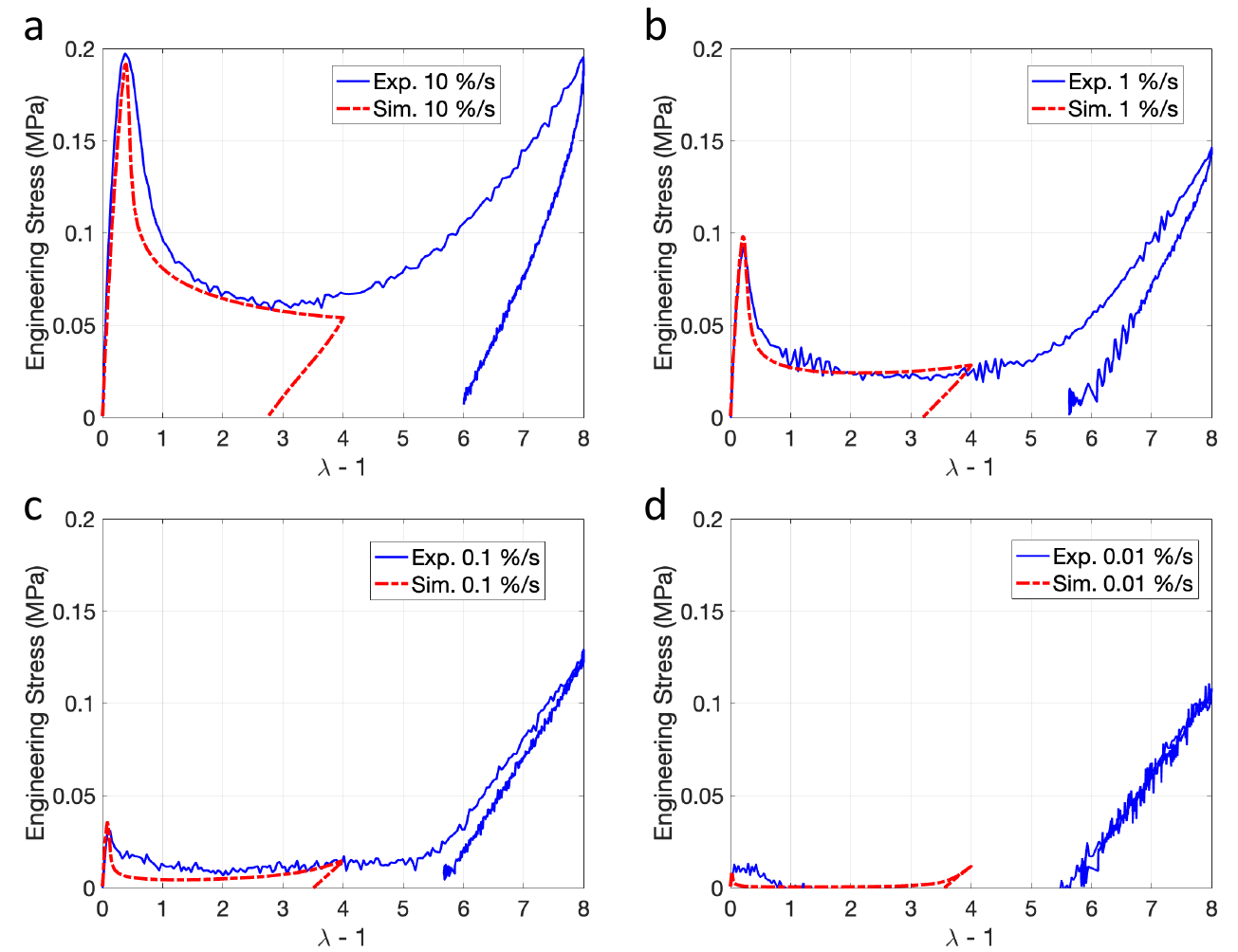}
    \caption{Material point simulations fitted to uniaxial tension responses of the perpendicular LCE with lower crosslinking density at strain rates of (a) 10\%/s, (b) 1\%/s, (c) 0.1\%/s, and (d) 0.01\%/s.}
    \label{fig:Fig_S3}
\end{figure}

\begin{figure}[H]
    \centering
    \includegraphics[width=0.77\textwidth]{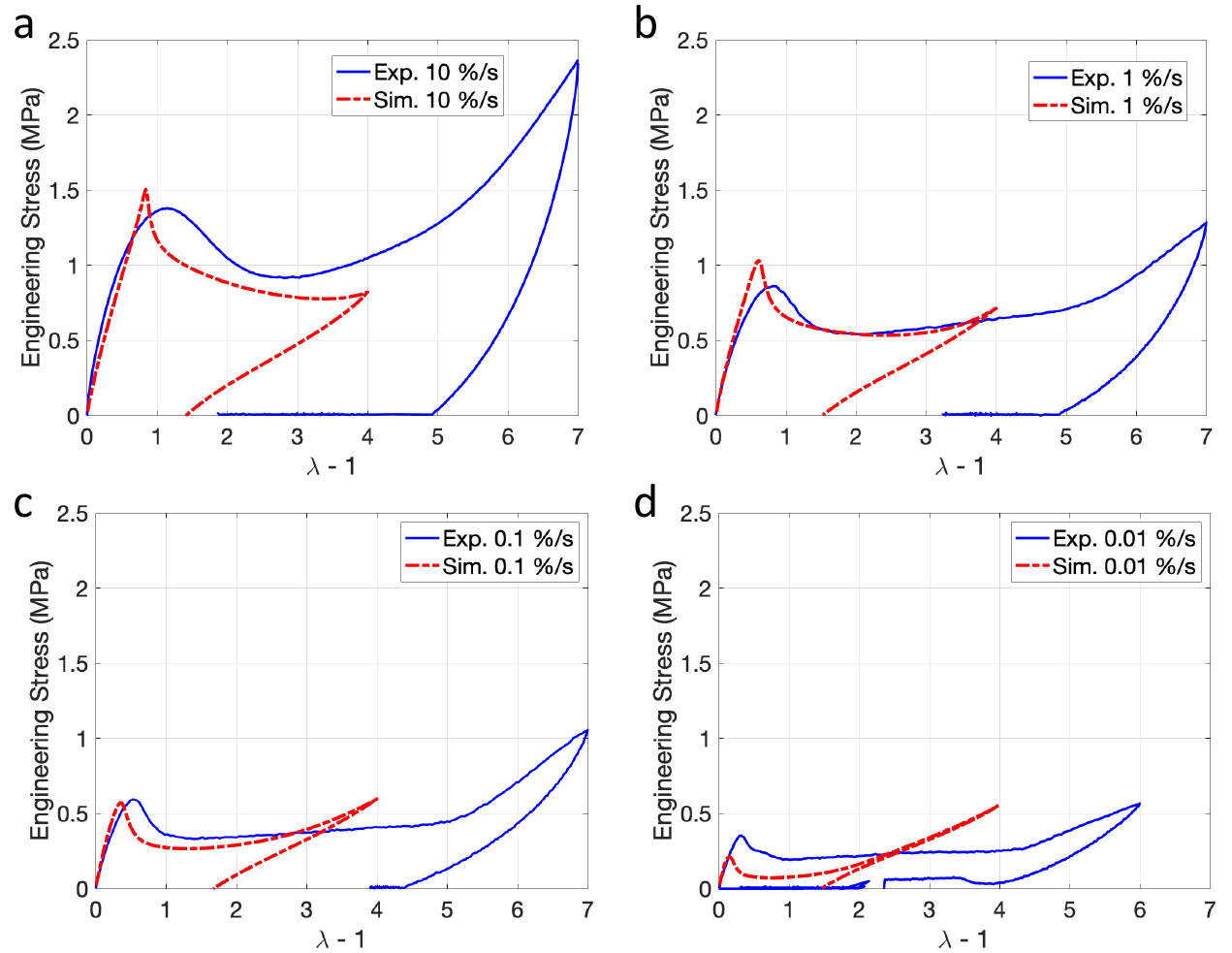}
    \caption{Material point simulations fitted to uniaxial tension responses of the perpendicular LCE with higher crosslinking density at strain rates of (a) 10\%/s, (b) 1\%/s, (c) 0.1\%/s, and (d) 0.01\%/s.}
    \label{fig:Fig_S4}
\end{figure}

Model parameters were calibrated by fitting the stress–strain response across all strain rates. For the lower-CLD LCE, the total instantaneous shear modulus $\mu^\mathrm{eq} + \mu^\mathrm{neq} = 0.25$~MPa was obtained from the experimental Young’s modulus at 1\%/s, assuming incompressibility. The director viscosity parameters $\eta_{D_0} = 11$~MPa$\cdot$s and $k_S = 0.1$ were fitted to capture the draw stress at 10\%/s, 1\%/s, and 0.1\%/s. The stiffening parameter $I_m$ was set large to suppress chain stretch effects. The network viscosity $\eta_N = 0.5$~MPa$\cdot$s was fitted to the residual strain after unloading. The order parameter $Q = 0.87$ controlled the soft plateau span, and the initial director angle $\theta_0 = 83^\circ$ was selected to match the experimental peak stress.

A similar procedure was applied for the higher-CLD LCE. The instantaneous shear modulus was $\mu^\mathrm{eq} + \mu^\mathrm{neq} = 0.8$~MPa. Director viscosity parameters $\eta_{D_0} = 1000$~MPa$\cdot$s and $k_S = 0.6$ were fitted to the draw stress at various rates. A much larger $\eta_N = 700$~MPa$\cdot$s reflected the denser network. The order parameter $Q = 0.62$ and director angle $\theta_0 = 80^\circ$ were calibrated to match the soft plateau and peak stress, respectively. All fitted parameters are summarized in Tables~\ref{tab:Model_SoftPerpLCE} and~\ref{tab:Model_StiffPerpLCE}.

Figures~\ref{fig:Fig_S3} and~\ref{fig:Fig_S4} compare simulation results to experimental curves. For the lower-CLD LCE, the model accurately captured the stress–strain response at all rates (Figure~\ref{fig:Fig_S3}). For the higher-CLD LCE (Figure~\ref{fig:Fig_S4}), simulations predicted a sharper peak and more pronounced softening than experiments, but reproduced key features such as the stress plateau, residual strain, and strain rate sensitivity. These results support the model’s fidelity in capturing the large-strain viscoelastic behavior of LCEs.

\section{Validation of energy absorption computation during uniaxial tension simulations}
\label{app:EnergyValidation_Tension}
\setcounter{table}{0}
\setcounter{figure}{0}

\begin{figure}[H]
    \centering
    \includegraphics[width=0.95\textwidth]{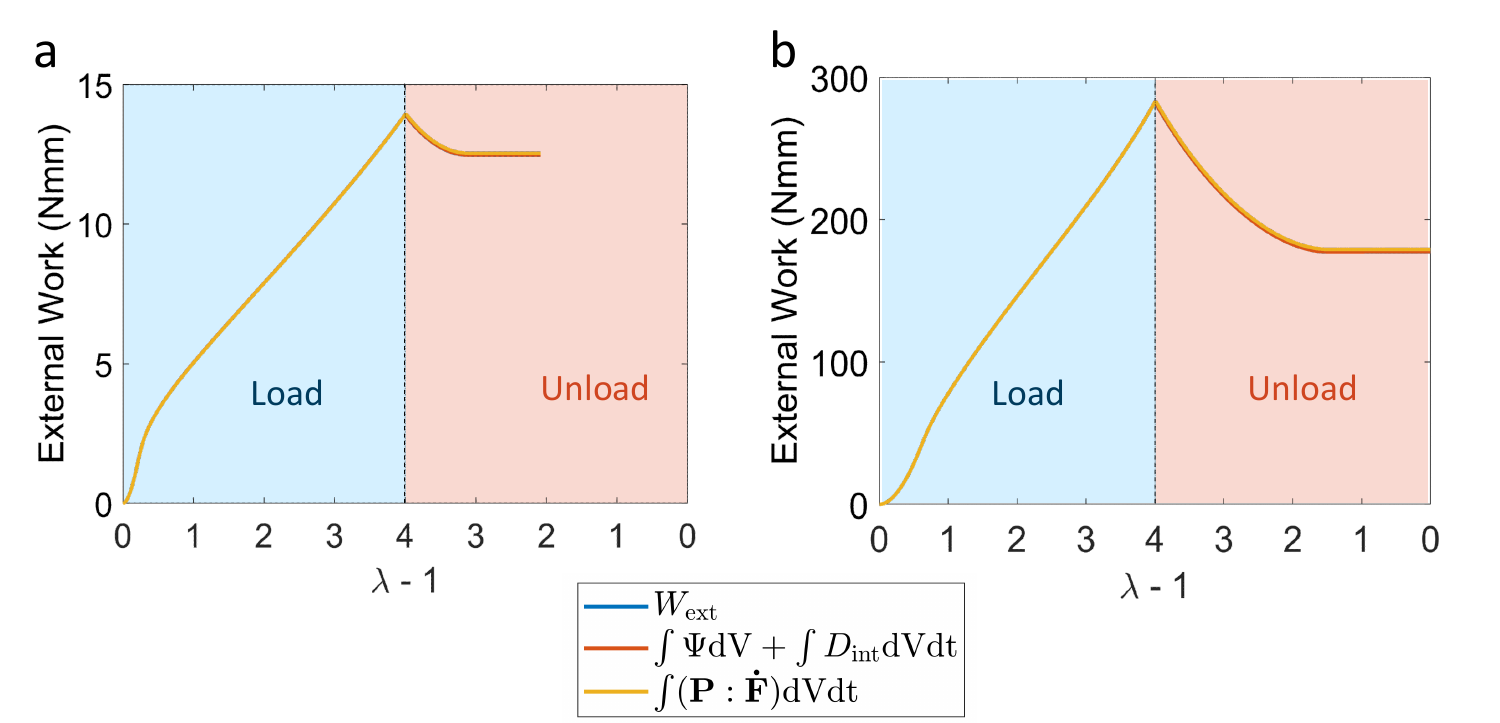}
    \caption{External work ($W_{\mathrm{ext}}$) computed by three methods from finite element simulations of uniaxial tensile load–unload cycles for perpendicular LCEs with (a) lower and (b) higher crosslinking density. Method 1: numerical integration of the force–displacement curve. Method 2: $W_{\mathrm{ext}} = \int \Psi \, \mathrm{dV} + \int D_{\mathrm{int}} \, \mathrm{dV} \, \mathrm{dt}$, where $\Psi$ is the Helmholtz free energy density and $D_{\mathrm{int}}$ is internal dissipation power per unit reference volume. Method 3: $W_{\mathrm{ext}} = \int (\mathbf{P} : \dot{\mathbf{F}}) \, \mathrm{dV} \, \mathrm{dt}$, where $\mathbf{P}$ is the first Piola–Kirchhoff stress and $\dot{\mathbf{F}}$ is the deformation rate.}
    \label{fig:Fig_S5}
\end{figure}

To ensure consistency in computing the external work during uniaxial tension simulations, we verified the results using three independent methods, as illustrated in Figure~\ref{fig:Fig_S5} for perpendicular LCEs with both crosslinking densities.

\section{Determining parameters for the finite-strain viscoelastic model}
\label{app:ParameterForTiltedBeams}
\setcounter{table}{0}
\setcounter{figure}{0}

The storage modulus master curves (Figure~\ref{fig:Fig_S0}a–b) were used to extract viscoelastic parameters for modeling the tilted LCE beams. Relaxation spectra $(\tau_k, \mu_k^{\mathrm{neq}})$ were obtained from the curves. For the polydomain LCE with higher crosslinking density, a fractional viscoelastic model was used to capture the experimental response with minimal fitting parameters~\citep{Haupt2000OnParameters}.

\begin{figure}[H]
    \centering     
    \includegraphics[width=0.35\textwidth]{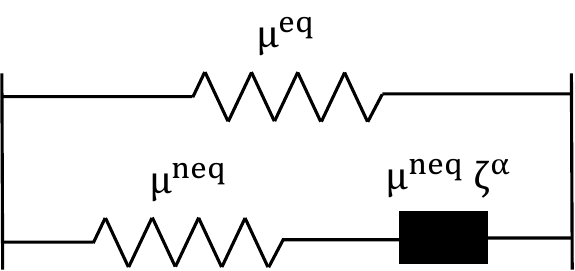}
    \caption{Fractional viscoelastic model with equilibrium and nonequilibrium shear moduli $\mu^{\mathrm{eq}}$, $\mu^{\mathrm{neq}}$, spectral breadth $\alpha$, and characteristic relaxation time $\zeta$. Reproduced from~\citep{Haupt2000OnParameters}, with permission from Elsevier.}
    \label{fig:Fig_S6}
\end{figure}

Following fractional calculus~\citep{Caputo1971LinearSolids, Koeller1984ApplicationsViscoelasticityb, Bagley1986OnBehavior}, the fractional derivative $\frac{d^\alpha f}{dt^\alpha}$ for $0 < \alpha < 1$ is defined as~\citep{Haupt2000OnParameters}:
\begin{equation}
    \frac{d^\alpha f}{dt^\alpha} = \frac{1}{\Gamma(1 - \alpha)} \int_0^t \frac{f'(s)}{(t - s)^\alpha} \, ds, \quad f(0) = 0,
    \label{eq:frac_derivative}
\end{equation}
where $\Gamma(\cdot)$ is the Gamma function. The fractional damping element~\citep{Lion1997OnElements, HSchiessel1995GeneralizedSolutions} relates stress to fractional strain rate:
\begin{equation}
    \frac{d^\alpha \epsilon}{dt^\alpha} = \frac{\sigma}{E \zeta^\alpha}.
\end{equation}
Here, $E$, $\zeta$, and $\alpha$ are non-negative parameters. The model recovers Hookean and Newtonian behavior as $\alpha \rightarrow 0$ and $1$, respectively (Figure~\ref{fig:Fig_S6}). Storage and loss moduli are given by~\citep{Haupt2000OnParameters, Nguyen2010ModelingPolymers}:
\begin{equation}
\begin{aligned}
    E_{frac}'(\omega) &= 3 \mu^{\mathrm{eq}} + \frac{2.7 \mu^{\mathrm{neq}}((\omega \zeta)^{2\alpha} + (\omega \zeta)^{\alpha} \cos(\alpha \pi / 2))}{1 + (\omega \zeta)^{2\alpha} + 2 (\omega \zeta)^{\alpha} \cos(\alpha \pi / 2)}, \\
    E_{frac}''(\omega) &= \frac{3 \mu^{\mathrm{neq}} (\omega \zeta)^{\alpha} \sin(\alpha \pi / 2)}{1 + (\omega \zeta)^{2\alpha} + 2 (\omega \zeta)^{\alpha} \cos(\alpha \pi / 2)}.
\end{aligned}
\end{equation}

\begin{figure}[H]
    \centering     
    \includegraphics[width=0.6\textwidth]{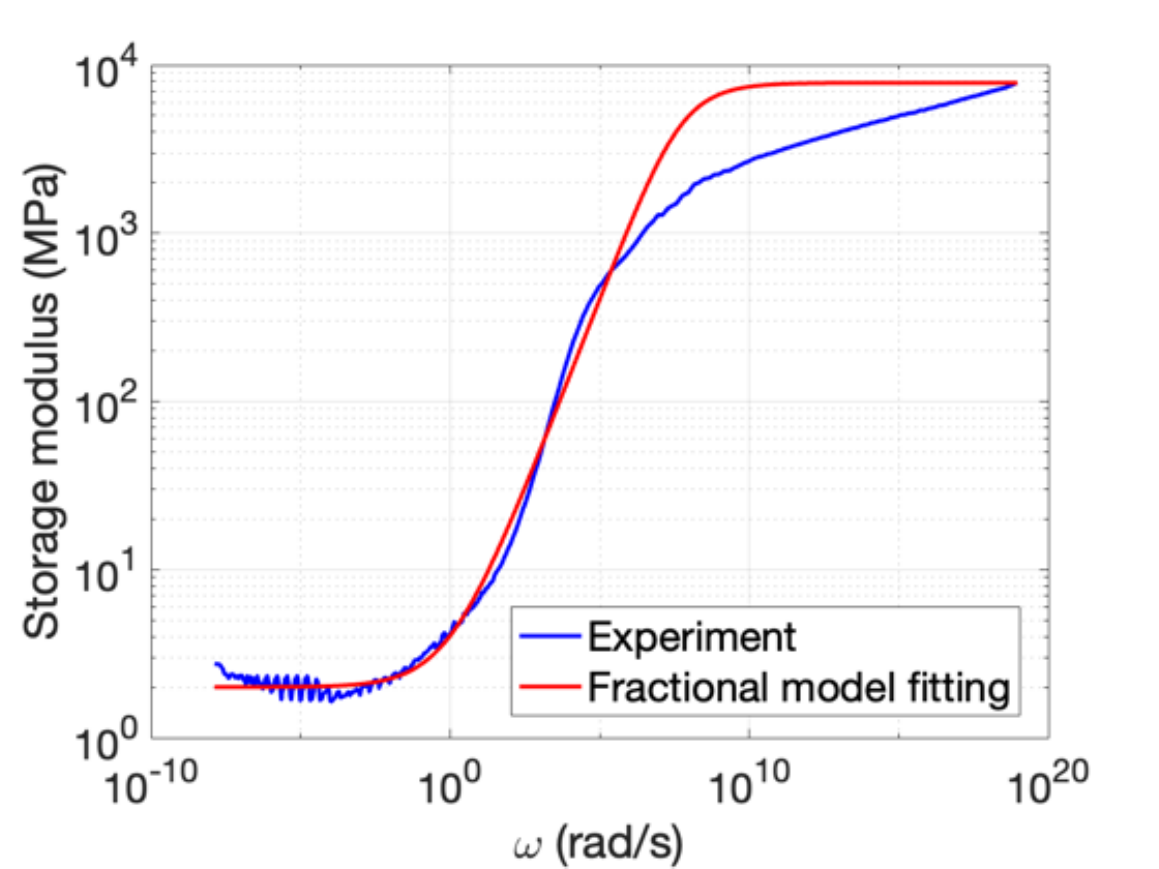}
    \caption{Storage modulus master curve fit using the fractional model at 25~$^\circ$C.}
    \label{fig:Fig_S7}
\end{figure}

The parameters $\mu^{\mathrm{eq}}$ and $\mu^{\mathrm{neq}}$ were estimated from the rubbery and glassy plateaus, assuming $\nu_g = 0.35$ and $\nu_r = 0.5$. The fractional exponent $\alpha$ controls the relaxation spectrum breadth, and $\zeta$ sets the characteristic time scale. Fitting yielded $\zeta = 3 \times 10^{-8}$~s and $\alpha = 0.46$ (Figure~\ref{fig:Fig_S7}). Deviations at high frequencies were considered negligible given the focus on moderate rates and lower frequencies.

The continuous relaxation spectrum is:
\begin{equation}
    h_{\mathrm{frac}}(\omega) = \frac{2.7 \mu^{\mathrm{neq}}(\omega \zeta)^{\alpha} \sin(\alpha \pi)}{\pi \omega \left[1 + (\omega \zeta)^{2\alpha} + 2 (\omega \zeta)^{\alpha} \cos(\alpha \pi)\right]}.
    \label{eq:frac_hw}
\end{equation}
Its cumulative form is:
\begin{equation}
    H_{\mathrm{frac}}(\omega) = \int_{0}^{\omega} h_{\mathrm{frac}}(z) \, dz,
\end{equation}
with analytic expression:
\begin{equation}
    H_{\mathrm{frac}}(\omega) = \frac{2.7 \mu^{\mathrm{neq}}}{\alpha \pi} \left[ \arctan\left(\frac{(\omega \zeta)^{\alpha} + \cos(\alpha \pi)}{\sin(\alpha \pi)}\right) - \pi \left(\frac{1}{2} - \alpha\right) \right].
    \label{eq:frac_Hw}
\end{equation}

Discrete relaxation frequencies were defined logarithmically:
\begin{equation}
    \omega_k = \omega_{\text{min}} \left( \frac{\omega_{\text{max}}}{\omega_{\text{min}}} \right)^{\frac{k-1}{N-1}}, \quad \tau_k = 1/\omega_k.
\end{equation}
Discrete moduli $\mu_i^{\mathrm{neq}}$ were calculated to approximate $H_{\mathrm{frac}}(\omega)$:
\begin{equation}
\begin{aligned}
    \mu_1^{\mathrm{neq}} &= \frac{1}{2.7} \cdot \frac{1}{2} \left[ H_{\mathrm{frac}}(\omega_1) + H_{\mathrm{frac}}(\omega_2) \right], \\
    \mu_k^{\mathrm{neq}} &= \frac{1}{2.7} \cdot \frac{1}{2} \left[ H_{\mathrm{frac}}(\omega_{k+1}) - H_{\mathrm{frac}}(\omega_{k-1}) \right], \quad 2 \leq k \leq N-1, \\
    \mu_N^{\mathrm{neq}} &= \mu^{\mathrm{neq}} - \sum_{k=1}^{N-1} \mu_k^{\mathrm{neq}}.
\end{aligned}
\end{equation}

\begin{figure}[H]
    \centering
    \includegraphics[width=\textwidth]{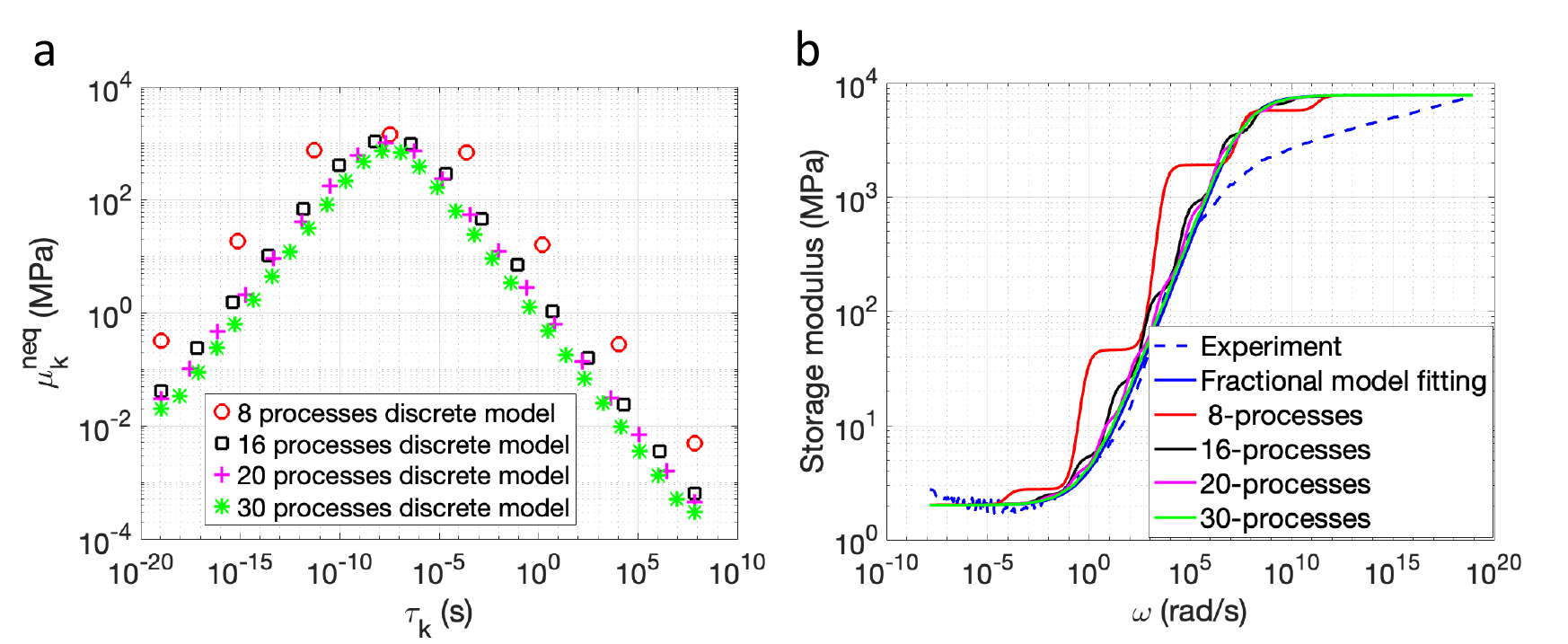}
    \caption{(a) Discrete relaxation spectra for different numbers of processes. (b) Reconstructed master curves of storage modulus for polydomain LCE with higher crosslinking density.}
    \label{fig:Fig_S8}
\end{figure}

The final model used $N = 20$ to balance accuracy and cost (Figure~\ref{fig:Fig_S8}). Final parameters for the polydomain LCE with higher crosslinking density are summarized in Table~\ref{tab:Parameters_StiffPD}, including relaxation times $\tau_i^{\mathrm{neq}}$, shear moduli $\mu_i^{\mathrm{neq}}$, viscosities $\eta_i^{\mathrm{neq}}$, and the normalized Prony series terms dimensionless shear (deviatoric) $g_i$ and bulk (volumetric) moduli $k_i$ for Abaqus \citep{dassaultsystemesSIMULIAUserAssistance2023}.

Normalized moduli $g_i$ are defined as:
\begin{equation}
g_i = \frac{\mu_i^{\mathrm{neq}}}{\mu_0} = \frac{\mu_i^{\mathrm{neq}}}{\sum_{i=1}^{N} \mu_i^{\mathrm{neq}} + \mu^{\mathrm{eq}}},
\end{equation}
with $\mu_0$ as the instantaneous shear modulus. Bulk relaxation was neglected by setting $k_i = 0$. The hyperelastic baseline used $C_{10} = \mu^{\mathrm{eq}}/2 = 0.333$ MPa and $D_1 = 2/\kappa^{\mathrm{eq}} = 2.29 \times 10^{-4}$.

\begin{figure}[H]
    \centering     
    \includegraphics[width=\textwidth]{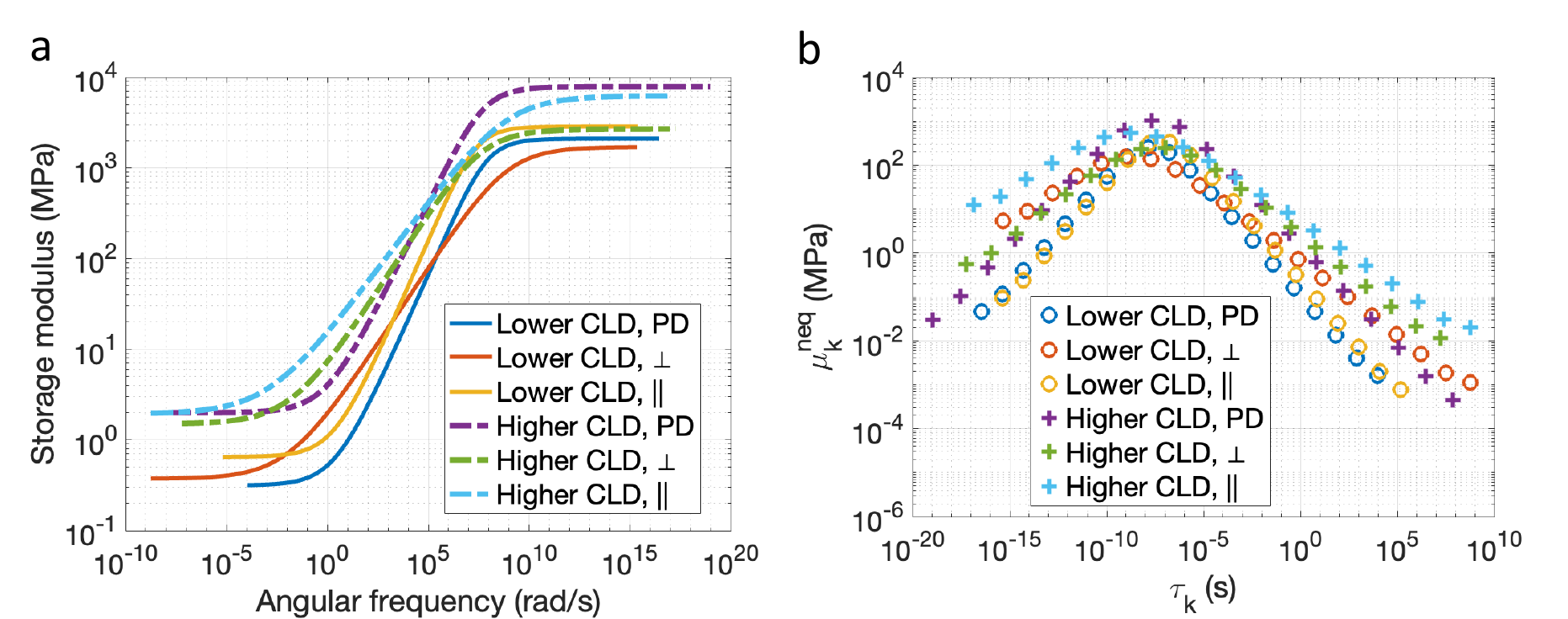}
    \caption{(a) Fractional model fits to storage modulus master curves for six LCE types. (b) Corresponding discrete relaxation spectra.}
    \label{fig:Fig_S9}
\end{figure}

The same fitting procedure was applied to all six LCE types. Their fitted storage modulus curves and discrete spectra are shown in Figure~\ref{fig:Fig_S9}.

\begin{table}[H]
    \centering
    \caption{Material parameters for modeling the polydomain LCE with higher crosslinking density.}
    \begin{tabular}{@{}cccccc@{}}
        \toprule     \shortstack{\textbf{Non-equilibrium} \\ \textbf{branches}} & $\bm{\tau_i^{\mathrm{neq}}}$ (s) & $\bm{\mu_i^{\mathrm{neq}}} $ (MPa)& $\bm{\eta_i^{\mathrm{neq}}}$ (MPa$\cdot$s) & $\bm{g_i}$ & $\bm{k_i}$ \\ 
        \midrule
        1  & 7.45$\times 10^7$   & 4.51$\times 10^{-4}$  & 3.36$\times 10^4$  & 1.55$\times 10^{-7}$  & 0   \\
        2  & 2.88$\times 10^6$   & 1.56$\times 10^{-3}$  & 4.51$\times 10^3$  & 5.38$\times 10^{-7}$  & 0   \\
        3  & 1.12$\times 10^5$   & 6.98$\times 10^{-3}$  & 7.79$\times 10^2$  & 2.40$\times 10^{-6}$  & 0   \\
        4  & 4.32$\times 10^3$   & 3.11$\times 10^{-2}$  & 1.35$\times 10^2$  & 1.07$\times 10^{-5}$  & 0  \\
        5  & 1.67$\times 10^2$   & 1.39$\times 10^{-1}$  & 23.2  & 4.78$\times 10^{-5}$ & 0  \\
        6  & 6.47   & 6.20$\times 10^{-1}$  & 4.01  & 2.13$\times 10^{-4}$  & 0   \\
        7  & 2.51$\times 10^{-1}$   & 2.77  & 6.93$\times 10^{-1}$  & 9.52$\times 10^{-4}$  & 0  \\
        8  & 9.70$\times 10^{-3}$   & 12.3  & 1.20$\times 10^{-1}$  & 4.24$\times 10^{-3}$  & 0   \\
        9  & 3.75$\times 10^{-4}$   & 54.7  & 2.05$\times 10^{-2}$  & 1.88$\times 10^{-2}$  & 0  \\
        10 & 1.45$\times 10^{-5}$   & 2.33$\times 10^2$  & 3.38$\times 10^{-3}$  & 8.01$\times 10^{-2}$  & 0  \\
        11 & 5.62$\times 10^{-7}$   & 7.34$\times 10^2$  & 4.13$\times 10^{-4}$  & 2.53$\times 10^{-1}$  & 0  \\
        12 & 2.18$\times 10^{-8}$   & 1.02$\times 10^3$  & 2.22$\times 10^{-5}$  & 3.51$\times 10^{-1}$  & 0   \\
        13 & 8.42$\times 10^{-10}$   & 6.18$\times 10^2$  & 5.21$\times 10^{-7}$  & 2.13$\times 10^{-1}$  & 0  \\
        14 & 3.26$\times 10^{-11}$   & 1.76$\times 10^2$  & 5.75$\times 10^{-9}$  & 6.07$\times 10^{-2}$  & 0  \\
        15 & 1.26$\times 10^{-12}$   & 40.8  & 5.15$\times 10^{-11}$  & 1.40$\times 10^{-2}$  & 0  \\
        16 & 4.88$\times 10^{-14}$   & 9.18  & 4.49$\times 10^{-13}$  & 3.16$\times 10^{-3}$  & 0  \\
        17 & 1.89$\times 10^{-15}$   & 2.06  & 3.89$\times 10^{-15}$  & 7.09$\times 10^{-4}$  & 0  \\
        18 & 7.32$\times 10^{-17}$   & 4.62$\times 10^{-1}$  & 3.38$\times 10^{-17}$  & 1.59$\times 10^{-4}$  & 0   \\
        19 & 2.83$\times 10^{-18}$   & 1.03$\times 10^{-1}$  & 2.93$\times 10^{-19}$  & 3.56$\times 10^{-5}$  & 0   \\
        20 & 1.10$\times 10^{-19}$   & 2.99$\times 10^{-2}$  & 3.27$\times 10^{-21}$  & 1.03$\times 10^{-5}$  & 0  \\ 
        \bottomrule
    \end{tabular}
    \label{tab:Parameters_StiffPD}
\end{table}

\section{Finite element simulations of unit cell structures: effect of tilted LCE beam properties}
\label{app:FEM_Structures}
Beyond the parametric studies on horizontal LCE bar thickness and material properties, we further examined how the viscoelastic behavior of the tilted LCE beams influences energy absorption in the hexagonal structure. This analysis aimed to clarify how beam material selection affects the balance between beam bending and bar stretching.

Prior to simulation, viscoelastic parameters for all six LCE types were extracted from DMA master curves and fitted using the fractional viscoelastic model described in Appendix~\ref{app:ParameterForTiltedBeams}. The corresponding discrete relaxation spectra for $N = 20$ non-equilibrium processes are shown in Figure~\ref{fig:Fig_S9}.

\begin{figure}[H]
    \centering     
    \includegraphics[width=\textwidth]{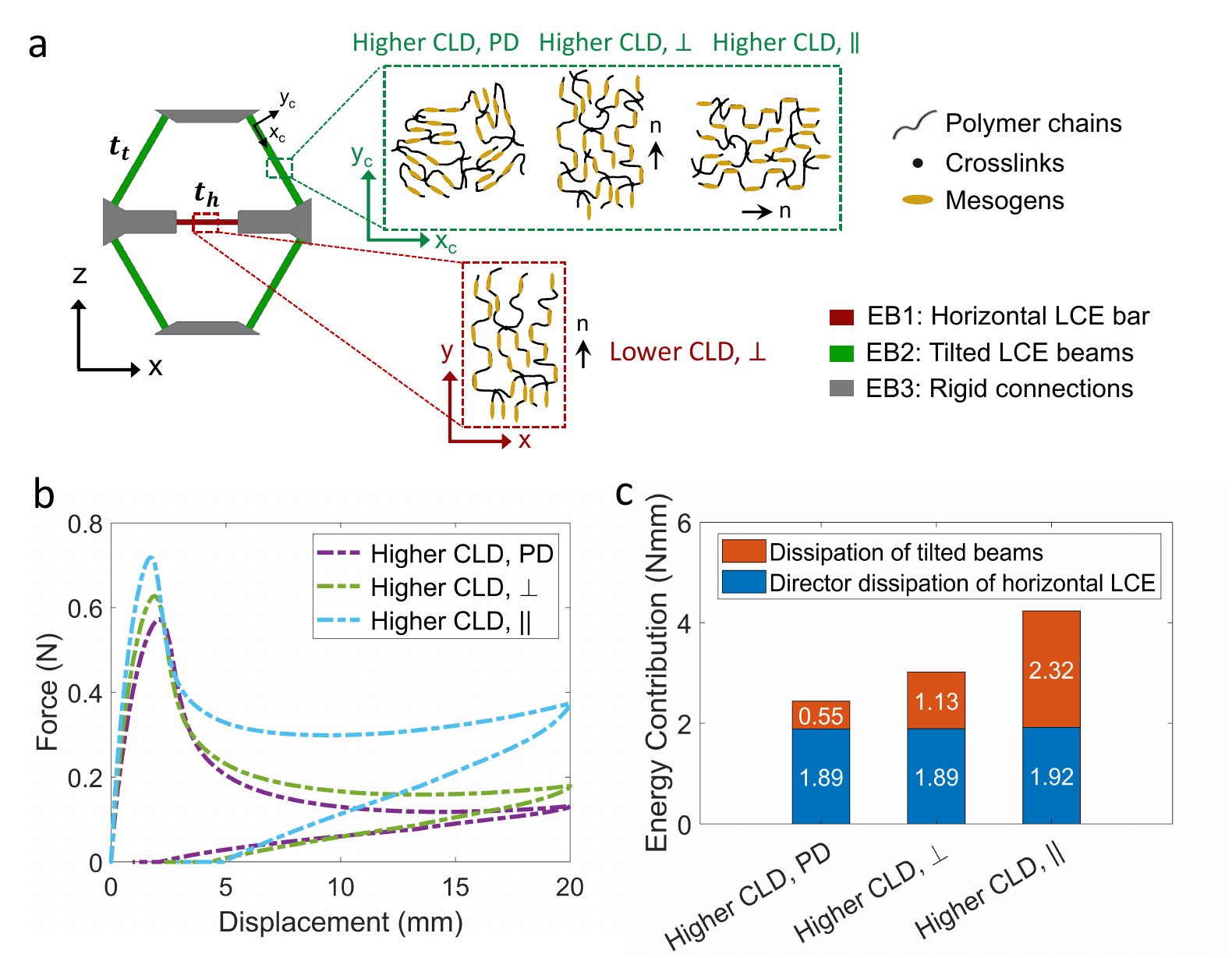}
    \caption{(a) Schematic of the unit cell structure with a horizontal LCE bar. The tilted beams are made from LCEs with higher crosslinking density (CLD): polydomain (PD), perpendicular ($\perp$), or parallel ($\parallel$). The horizontal bar is perpendicular LCE with lower CLD. 
    (b) Simulated force–displacement curves using different tilted beam materials. 
    (c) Corresponding energy contributions from beam bending and director rotation in the horizontal bar. Tilted beam and horizontal bar thicknesses are fixed at $t_t = 0.93$~mm and $t_h = 0.64$~mm, respectively.}
    \label{fig:Fig_S10}
\end{figure}

We performed a parametric study using three LCE types—polydomain, perpendicular, and parallel—each with higher crosslinking density, as the material for the tilted beams. The horizontal LCE bar was fixed as a perpendicular LCE with lower crosslinking density (Figure~\ref{fig:Fig_S10}a). Figures~\ref{fig:Fig_S10}b–c show the simulation results. All cases exhibited stretch-dominant behavior, characterized by an initial force rise, a peak, and a sudden drop followed by a plateau, as previously observed in Figures~\ref{fig:Fig_6} and~\ref{fig:Fig_7}b, d, f. 
Structural stiffness increased from polydomain to perpendicular to parallel LCEs. Dissipation from director rotation in the horizontal bar remained nearly constant, while viscoelastic bending dissipation increased with beam stiffness, from lowest in polydomain to highest in parallel. These results indicate that, even with a fixed horizontal bar and stretch-dominant behavior, tilted beam properties significantly affect both force response and energy absorption.

\end{appendices}

%%       \citet{<label>} ==> Lamport [21]
%%       \citep{<label>} ==> [21]
%%




\bibliographystyle{elsarticle-num-names} 
\bibliography{references}  % assuming your .bib file is named references.bib

\end{document}